\documentclass[aps,pra,twocolumn,superscriptaddress,showpacs,preprintnumbers,amsmath,amssymb,nofootinbib]{revtex4}
\usepackage{graphicx}
\usepackage{dcolumn}
\usepackage{bm}

\usepackage[usenames]{color}

\begin{document}

\title{Lee-Yang cluster expansion approach to the BCS-BEC crossover: BCS and BEC limits}

\author{Naoyuki~Sakumichi}
\affiliation{Theoretical Research Division, Nishina Center, RIKEN, Wako, Saitama 351-0198, Japan}%
\author{Yusuke~Nishida}
\affiliation{Department of Physics, Tokyo Institute of Technology, Ookayama, Meguro, Tokyo 152-8551, Japan}
\author{Masahito~Ueda}
\affiliation{Department of Physics, The University of Tokyo, Hongo, Bunkyo, Tokyo 113-0033, Japan}%

\date{\today}
\begin{abstract} 
It is shown that a cluster expansion technique, which is usually applied in the high-temperature regime to calcutate virial coefficients, can be applied to evaluate the superfluid transition temperature of the BCS-BEC crossover \`a la Lee and Yang.
The transition temperature is identified with the emergence of the singularity in the sum of a certain infinite series of cluster functions.
In the weak-coupling limit, we reproduce the Thouless criterion and the number equation of Nozi\`eres and Schmitt-Rink, and hence the transition temperature of the BCS theory.
In the strong-coupling limit, we reproduce the transition temperature of Bose-Einstein condensation of non-interacting tightly bound dimers.
\end{abstract}


\maketitle



\section{INTRODUCTION}
This paper concerns the application of a cluster expansion technique to the Bardeen-Cooper-Schrieffer (BCS)-Bose-Einstein condensation (BEC) crossover of a dilute gas of two-component Fermi particles with zero-range interaction,
as has been realized using ultracold atomic gases \cite{EaglesLeggett, GPS08, Zwerger12}.
Here, the zero-range means that the range of the inter-particle potential $r_0$ is much shorter than the inverse Fermi wavenumber $k_F^{-1}$, the $s$-wave scattering length $a$, and the thermal de Broglie length $\lambda := (2\pi \hbar^2 /m k_B T)^{1/2}$, i.e., $r_0 \ll k_F^{-1}, \lambda, \left| a \right|$, where $m$ is the mass of a particle and $T$ is the temperature.
It is widely held that this system possesses a universal property that the equation of state depends only on $a$, $k_F$, and $\lambda$,
and that the phase diagram is characterized by the temperature $T/T_F$ and the dimensionless interaction parameter $(k_F a)^{-1}$, where $T_F$ is the Fermi temperature.
Therefore, this system provides a simple universal model for understanding various degenerate Fermi systems such as a quark-gluon plasma \cite{Shuryak04}, neutron stars \cite{Bertsch00}, excitons \cite{LEKMSS04}, and high-$T_c$ superconductors \cite{CSTL05}.
There are many theoretical approaches to the phase transition based on unbiased quantum Monte Carlo techniques \cite{QMC}, 
a functional renormalization group method \cite{FRG}, and Feynman diagrammatic techniques \cite{NSR85, MRE93}.
%

\begin{figure}[b]
\begin{picture}(245,110)
\put(25,-5){\includegraphics[width=200pt]{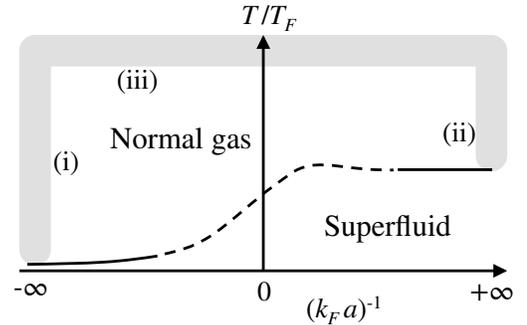}}
\end{picture}
\caption{
Schematic phase diagram of the BCS-BEC crossover 
as a function of the temperature $T/T_F$ 
and the inverse scattering length $(k_F a)^{-1}$.
In this paper,
we establish the theory, which correctly captures the physics in (i) the weak- and (ii) strong-coupling limits at and above the transition temperature and (iii) in the high-temperature regime for any $(k_Fa)^{-1}$
based on the Lee-Yang cluster expansion method.
}
\label{fig:BCS-BEC_crossover}
\end{figure}

In this paper, we provide a new approach to the phase transition of this system based on cluster expansion in terms of the fugacity $z := \exp [\mu/ (k_B T)]$ \cite{Huang}, where $\mu$ is the chemical potential.
Although the cluster-expansion approach is usually applied at high temperatures \cite{HM04, LHD10, KS11, L11, RDB12, LMNR12, AZFJ12, Liu12, CW12},
we show that it can be applied to evaluate the transition temperature  for the onset of quantum condensation of the BCS-BEC crossover \`a la Lee and Yang.
The method of Lee and Yang \cite{LY58-I, LY60-IV, Mo61, dD62, OU06, In09, SKU12} allows systematic evaluation of the higher-order cluster integrals in terms of the cluster functions (Ursell functions) which are in terms of the quantum-mechanical problem with Boltzmann statistics.
We show that the proposed theory correctly captures the physics in the weak-coupling, strong-coupling, and high-temperature regimes (see Fig.~\ref{fig:BCS-BEC_crossover}).
We can evaluate the transition temperature by analyzing an emergence of the singularity of an infinite series of cluster functions.
We identify a certain infinite series of cluster functions, the sum of which has the following three properties: 
(i) in the weak-coupling limit (so-called BCS limit), 
it gives the Thouless criterion \cite{T60} and hence the transition temperature of the BCS theory, and the number equation of the BCS-BEC crossover theory of Nozi\`eres and Schmitt-Rink (NSR) \cite{NSR85, MRE93};
(ii) in the strong-coupling limit (so-called BEC limit), 
it reproduces the thermodynamic function of non-interacting tightly bound dimers; 
(iii) in the high-temperature regime $T \gtrsim T_F$ for an arbitrary $s$-wave scattering length $(k_Fa)^{-1}$, it reproduces the exact second virial coefficient and thus Tan's contact \cite{Tan08-a, Tan08-b, Tan08-c} up to the same order.
All of these suggest that our theory provides a good starting point to describe the entire BCS-BEC crossover.

We mention here recent cluster-expansion studies of evaluating the low-order cluster integrals (or the virial coefficients) \cite{HM04, LHD10, KS11, L11, RDB12, LMNR12, AZFJ12, Liu12, CW12}.
These studies have an advantage over a Feynman diagrammatic technique because the fugacity $z$ is a controllable small parameter when the system is dilute and at high temperatures.
In addition, the equation of state of a homogeneous Fermi gas in the unitary limit ($(k_F a)^{-1}=0$) has recently been measured by using a two-component mixture of $^6$Li atoms \cite{HNUM10, NNJCS10, KSCZ11}.
These experiments demonstrate that the low-order (third-order) cluster expansion quantitatively describes the equation of state down to temperatures as low as the Fermi temperature ($T \gtrsim T_F$), which corresponds to $z<1$ \cite{LHD10, NNJCS10, KSCZ11, Liu12}.
Although the low-order cluster expansion well describes the region $z<1$, 
a perturbative cluster-expansion calculation cannot describe the phase transition,
because a thermodynamic function, which is obtained by a low-order-cluster-expansion calculation, is a polynomial of $z$ and thus has no singularity corresponding to the phase transition point.
In fact, the fugacity at the superfluid phase transition $z_c$ is greater than $10$ \cite{NNJCS10, KSCZ11}.
Therefore, to analyze the phase transition, we should take into account higher-order terms appropriately.
In the present study, we demonstrate that a cluster-expansion method can be used to evaluate the transition temperatures of the BCS-BEC crossover, at least in the weak- and strong-coupling limits.

This paper is organized as follows.
In Sec.~II, we formulate the cluster-expansion method for the case of (pseudo) spin-$1/2$ Fermi systems.
In Sec.~III, we describe our model and apply the cluster-expansion method of Lee and Yang to this model.
We calculate the second cluster integral for an arbitrary $s$-wave scattering length, which reproduce the standard Beth-Uhlenbeck result \cite{HM04} and Tan's contact up to the same order \cite{YBB09, HLD11}.
Our method can thus treat effects of quantum-mechanical scattering and a bound state at least at the level of the second cluster integral.
In Sec.~IV,
we apply the Lee-Yang method to evaluate the transition temperature for the onset of quantum condensation of the BCS-BEC crossover.
In the weak-coupling limit, we reproduce the Thouless criterion and the number equation of the BCS-BEC crossover theory by NSR.
In the strong-coupling limit, we reproduce the BEC of dimers below a transition temperature.
In Sec.~V, we summarize the main results of this paper.
The details of the Lee-Yang method and the proofs of several formulas are described in Appendices to avoid digressing from the main subject.


\section{Lee-Yang cluster expansion method}
\label{sec:LYCE}

In this section, we describe the cluster expansion of the equation of state \cite{Huang} and that of Tan's contact \cite{YBB09, HLD11}, and define the cluster functions $U^{(N_\uparrow, N_\downarrow)}$.
By using the method of Lee and Yang \cite{LY58-I, LY60-IV, dD62, SKU12}, these cluster expansions can be expressed in terms of the primary or contracted $\zeta$-graphs ($\zeta =0, 1, 2, \dots$) which are computed from the cluster functions. 
The definitions of the primary and contracted $\zeta$-graphs and the details of the method of Lee and Yang are described in Appendix~\ref{App:LYCE}.

\subsection{Cluster expansion of equation of state and Tan's contact}

We consider a system of two-component (or pseudo-spin-$1/2$) fermions with the same mass $m$ and confined in a finite volume $V=L^3$ with periodic boundary conditions.
When $N_\uparrow$ particles have spin $\uparrow$ and $N_\downarrow$
particles  have spin $\downarrow$,
the partition function is
\begin{equation}
Z_V^{(N_\uparrow, N_\downarrow)}
 := \sum_{i} e^{- \beta E_i},
\label{eq:def:PatritionFn}
\end{equation}
where $\beta=1/k_B T$ is the inverse temperature and $E_i$ is the energy eigenvalue of the Hamiltonian $H^{(N_\uparrow, N_\downarrow)}$.
%
We assume that the chemical potentials $\mu$ is independent of the (pseudo-)spin states $\sigma=\uparrow,\downarrow$.
The grand partition function is
\begin{equation}
\Xi_V := \sum_{N_\uparrow=0}^{\infty} \sum_{N_\downarrow=0}^{\infty} z^{N_\uparrow + N_\downarrow} Z_V^{(N_\uparrow, N_\downarrow)},
\label{eq:def:GPF}
\end{equation}
where $z=e^{\beta \mu}$ is the fugacity and we define $Z_V^{(0, 0)} :=1$.
According to the principles of statistical mechanics,
the equilibrium pressure $p$, the particle-number density $\rho = \rho_\uparrow + \rho_\downarrow$, 
and the energy per unit volume $\varepsilon$ of the system are given by
\begin{align}
& \beta p = \lim_{V\to\infty} \frac{1}{V} \ln \Xi_V,
\label{eq:pressure} \\
& \rho = \lim_{V\to\infty}
    \frac{1}{V} \, z \, \frac{\partial}{\partial z} \ln \Xi_V,
\label{eq:densitiy}
\end{align}
and
\begin{equation}
 \varepsilon  = - \lim_{V\to\infty}
    \frac{1}{V} \,  \frac{\partial}{\partial \beta}
    \ln \Xi_V.
\label{eq:energy-densitiy}
\end{equation}
By eliminating $z$ in Eqs.~(\ref{eq:pressure}) and (\ref{eq:densitiy}), we obtain the equation of state \cite{Huang}. 

We define the thermal de Broglie length by $\lambda := (2\pi \beta \hbar^2/m)^{1/2}$
and expand $\lambda^3 \beta p$ and $\lambda^3 \rho$ in terms of the fugacity $z$ as
\begin{align}
 \lambda^3 \beta p & = \sum_{n=1}^{\infty} b_n z^n,
\label{eq:pressure_Cluster_Expansion} \\
 \lambda^3 \rho & = \sum_{n=1}^{\infty} n b_n z^n.
\label{eq:density_Cluster_Expansion}
\end{align}
The set of Eqs.~(\ref{eq:pressure_Cluster_Expansion}) and (\ref{eq:density_Cluster_Expansion}) gives the cluster expansion of the equation of state.
From the knowledge of the cluster integrals up to the $l$-th order $b_1, \dots, b_l$,
we can find the virial coefficients up to the same order \cite{Huang}.

For a non-interacting Fermi system, the Hamiltonian is 
$ H_{\rm ideal}^{(N_\uparrow, N_\downarrow)} = - \sum_{i=1}^{N_\uparrow + N_\downarrow}  \frac{\hbar^2}{2m} \nabla_{i}^2 $ 
and
the grand partition function is
\begin{equation}
\begin{split}
\ln \Xi_{\rm V, ideal}
& = 2 \sum_{\bold{k}} \ln \left[1+z e^{-\beta \hbar^2 k^2/(2m)} \right] \\
& = -2 \, \frac{V}{\lambda^3} \operatorname{Li}_{\frac{5}{2}} (-z).
\end{split}
\end{equation}
Here $\operatorname{Li}_l(x):=\sum_{n=1}^\infty x^n/n^{l}$ is the polylogarithm.
Then, the equilibrium pressure $p_{\rm ideal}$ is given by
\begin{equation}
 \lambda^3 \beta p_{\rm ideal} 
= - 2 \operatorname{Li}_{\frac{5}{2}} (-z) .
\label{eq:Cluster_Expansion-ideal}
\end{equation}
Therefore, we obtain
\begin{equation}
 \lambda^3 \beta \Delta p = \sum_{n=2}^{\infty} \Delta b_n z^n,
\label{eq:Cluster_Expansion_Delta_p}
\end{equation}
where $\Delta p := p - p_{\rm ideal}$ and $\Delta b_n := b_n - 2 (-1)^{n+1}/n^{5/2}$.
From Eqs.~(\ref{eq:def:GPF}), (\ref{eq:pressure}), and (\ref{eq:pressure_Cluster_Expansion}), we find $b_1=2$ and $\Delta b_1 = 0$. 
The remaining problem is to calculate $\Delta b_n$ ($n \geq 2$).

In Sec.~\ref{sec:Crossover} and \ref{sec:PT},
we consider a zero-range interaction which is characterized by the $s$-wave scattering length $a$.
Then, $b_n$ and $\Delta b_n$ ($n \geq 2$) depend only on $\lambda/a$ from the dimensional analysis.
In addition the Fermi system with the zero-range interaction satisfies a set of universal exact relations known as Tan's relations \cite{Tan08-a, Tan08-b, Tan08-c}.
Let define $\left< \hat{n}_{\bold{k} \sigma}\right>$ be the statistical average of the number of particles with definite momentum $\bold{k}$ and spin $\sigma=\uparrow,\downarrow$ in a finite volume over the grand canonical ensemble. 
Here, $\left<n_{\bold{k} \sigma}\right>$ is normalized as $\lim_{V\to\infty} \sum_{\bold{k}, \sigma} \left< \hat{n}_{\bold{k} \sigma}\right>/V=\rho$ and can be shown to have the following $1/k^4$ tails at large momentum:
\begin{equation}
C \equiv \lim_{\lambda k\to\infty} k^4 \langle \hat{n}_{\bold{k} \uparrow} \rangle 
 = \lim_{\lambda k\to\infty} k^4 \langle \hat{n}_{\bold{k} \downarrow} \rangle ,
\label{eq:def:contact}
\end{equation}
where $C$ is so-called Tan's contact.
All of Tan's relations are governed by Tan's contact,
e.g., the pressure and the energy per unit volume are related as \cite{Tan08-b, Tan08-c}
\begin{equation}
p - \frac{2}{3} \varepsilon =  \frac{\hbar^2 C}{12\pi m a}.
\label{eq:pressure-relation}
\end{equation}
By using Eq.~(\ref{eq:pressure-relation}), the cluster expansion of Tan's contact \cite{YBB09, HLD11} can be obtained as follows.
From Eqs.~(\ref{eq:pressure}), (\ref{eq:energy-densitiy}), and (\ref{eq:pressure_Cluster_Expansion}),
we obtain
\begin{equation}
\begin{split}
\varepsilon
 = - \frac{\partial}{\partial \beta} (\beta p)
 = \frac{3}{2} p - \frac{1}{2a \lambda^2 \beta} \sum_{n=2}^{\infty} c_n z^n ,
\end{split}
\label{eq:energy-clusterexpansion}
\end{equation}
where we use the so-called contact coefficients \cite{HLD11}:
\begin{equation}
c_n := \frac{d b_n}{d (\lambda /a)} = \frac{d \Delta b_n}{d (\lambda /a)} .
\label{eq:def:ContactCoeff}
\end{equation}
Comparing Eqs.~(\ref{eq:pressure-relation}) and (\ref{eq:energy-clusterexpansion}),
we obtain the cluster expansion of Tan's contact:
\begin{equation}
C = \frac{8\pi^2}{\lambda^4} \sum_{n=2}^{\infty} c_n z^n .
\label{eq:CE_of_contact}
\end{equation}
In Sec.~\ref{sec:Crossover}-D, we show how to calculate the second-order contact coefficient $c_2$.

\begin{figure}
\begin{picture}(250,115)
\put(10,-5){\includegraphics[width=210pt]{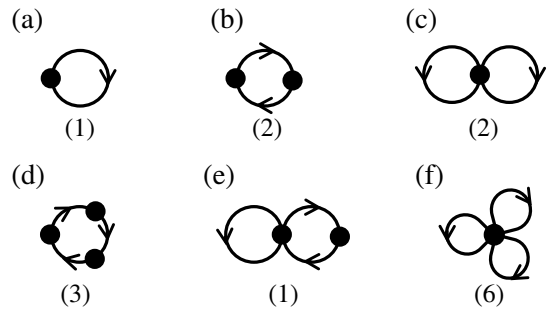}}
\end{picture}
\caption{
Primary $0$-graphs up to the third-order terms in fugacity $z$.
The graph in (a) contributes to the first-order term.
The graphs in (b) and (c) contribute to the second-order term.
The graphs in (d), (e), and (f) contribute to the third-order term.
The symmetry number is shown under each graph.
}
\label{fig:Primary_Graphs}
\end{figure}

\subsection{Cluster functions for a system obeying Boltzmann statistics}

By using the method of Lee and Yang \cite{LY58-I, LY60-IV},
the cluster expansion of the thermodynamic function for a system of particles obeying Fermi-Dirac statistics can be computed from the knowledge of the cluster functions for the same system obeying Boltzmann statistics.
We first introduce the functions
\begin{align}
& \langle \bold{r}_1',\dots,\bold{r}_N' | W^{(N_\uparrow, N_\downarrow)}  | 
  \bold{r}_1,\dots,\bold{r}_N \rangle \notag\\
& 
= 
\sum_{i}
 \psi_i \left( \bold{r}_1',\dots,\bold{r}_N' \right) 
 \psi_i^* \left( \bold{r}_1,\dots,\bold{r}_N \right)  e^{-\beta E_i},
\label{eq:def:W}
\end{align}
where $(\bold{r}_1,\dots,\bold{r}_N):=(\bold{r}_1,\dots,\bold{r}_{N_\uparrow};\bold{r}_{N_\uparrow+1},\dots,\bold{r}_N)$, 
$(\bold{r}_1',\dots,\bold{r}_N'):=(\bold{r}_1',\dots,\bold{r}_{N_\uparrow}';\bold{r}_{N_\uparrow+1}',\dots,\bold{r}_N')$,
particles $1,\dots,N_\uparrow$ have spin $\uparrow$, and
particles $N_\uparrow + 1,\dots,N_\uparrow + N_\downarrow = N$ have spin $\downarrow$.
Here, 
$\psi_{i}(\bold{r}_1,\dots,\bold{r}_N)$ and $E_i$ are the normalized eigenfunction and the corresponding eigenvalue of $H^{(N_\uparrow, N_\downarrow)}$ in the Hilbert space obeying Boltzmann statistics (i.e., the particles in this Hilbert space are distinguishable).
The summation in Eq.~(\ref{eq:def:W}) extends over all eigenvalues in this Hilbert space\footnote{
Thus, the sum in Eq.~(\ref{eq:def:W}) is different from that in Eq.~(\ref{eq:def:PatritionFn}) which runs over all eigenvalues in the Hilbert space obeying Fermi-Dirac statistics.
}.
The momentum representation of Eq.~(\ref{eq:def:W}) is defined by
\begin{equation}
\begin{split}
& \langle \bold{k}_1',\dots,\bold{k}_N' | W^{(N_\uparrow, N_\downarrow)}  | 
  \bold{k}_1,\dots,\bold{k}_N \rangle \\
&  = \frac{1}{V^{N}} 
     \int_{[0,L)^{3N}}
     \left( \prod_{\alpha=1}^N d^3\bold{r}_\alpha d^3\bold{r}_\alpha' \right)
   e^{i \sum_{\alpha=1}^N(\bold{k}'_\alpha\cdot\bold{r}'_\alpha -\bold{k}_\alpha\cdot\bold{r}_\alpha)}  \\
& \quad \times 
    \langle \bold{r}_1',\dots,\bold{r}_N' | W^{(N_\uparrow, N_\downarrow)}  | 
  \bold{r}_1,\dots,\bold{r}_N \rangle ,
\end{split}
\end{equation}
where $\bold{k}_i,\bold{k}'_i \in  (2\pi/L) \mathbb{Z}^3$.

We define the matrix elements of cluster functions $U^{(N_\uparrow, N_\downarrow)}$ in the momentum representation as
\begin{equation}
\begin{split}
     \langle \bold{k}'| W^{(1,0)}  | \bold{k} \rangle 
    & \equiv \langle \bold{k}'| U^{(1,0)}  | \bold{k} \rangle  
    = \delta_{\bold{k},\bold{k}'} e^{-\beta \bold{k}^2/(2m)},\\
     \langle \bold{k}'| W^{(0,1)}  | \bold{k} \rangle 
    & \equiv \langle \bold{k}'| U^{(0,1)}  | \bold{k} \rangle  
    = \delta_{\bold{k},\bold{k}'} e^{-\beta \bold{k}^2/(2m)},\\
    \langle 1', 2' | W^{(2,0)} | 1, 2 \rangle
    & \equiv  \langle 1', 2' | U^{(2,0)} | 1, 2 \rangle \\
    & \quad +  \langle 1' | U^{(1,0)} | 1 \rangle \, \langle 2' | U^{(1,0)} | 2 \rangle,\\
    \langle 1'; 2' | W^{(1,1)} | 1; 2 \rangle
    & \equiv  \langle 1'; 2' | U^{(1,1)} | 1; 2 \rangle \\
    & \quad +  \langle 1' | U^{(1,0)} | 1 \rangle \, \langle 2' | U^{(0,1)} | 2 \rangle,\\
    \langle 1', 2' | W^{(0,2)} | 1, 2 \rangle
    & \equiv  \langle 1', 2' | U^{(0,2)} | 1, 2 \rangle \\
    & \quad +  \langle 1' | U^{(0,1)} | 1 \rangle \, \langle 2' | U^{(0,1)} | 2 \rangle,\\
      \langle 1', 2' , 3' | W^{(3,0)} |  1, &2, 3 \rangle \\
      \equiv \langle 1',  2' , 3' |  U^{(3,0)} & | 1, 2 , 3 \rangle  \\
      +  \langle 1' |  U^{(1,0)}   | 1 \rangle \langle & 2', 3' | U^{(2,0)} | 2, 3 \rangle \\
      + \langle 2' |  U^{(1,0)}  | 2 \rangle \langle &  3', 1' | U^{(2,0)} | 3, 1 \rangle \\
      + \langle 3' |  U^{(1,0)}  | 3 \rangle \langle & 1', 2' | U^{(2,0)} | 1, 2 \rangle \\
      +\langle 1' |  U^{(1,0)}  | 1 \rangle \langle & 2' | U^{(1,0)} | 2 \rangle \langle 3' | U^{(1,0)} | 3 \rangle , 
 {\rm etc.}
\label{eq:def:U}
\end{split}
\end{equation}
Here,
$1 := \bold{k}_1$,
and
$1' := \bold{k}_{1}'$,
etc.
The two-particle cluster function $\langle 1'; 2' | U^{(1,1)} | 1; 2 \rangle$ is explicitly calculated in Sec.~III-B and Appendix~\ref{app:U2_pseudo}.

\begin{figure}
\begin{picture}(250,150)
\put(0,-3){\includegraphics[width=245pt]{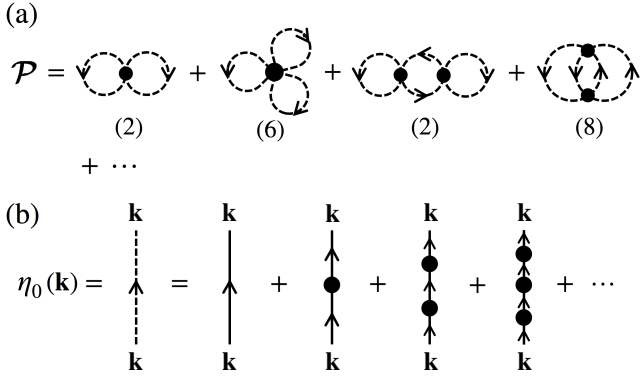}}
\end{picture}
\caption{
(a) Sum over all different contracted $0$-graphs (dotted curves).
The symmetry number is shown under each graph.
(b) Expression of $\eta_0 (\bold{k})$
as the sum over different primary $1$-graphs (solid lines).}
\label{fig:Xi}
\end{figure}

\subsection{Thermodynamic function and reduced density matrices in terms of Lee-Yang $\zeta$-graphs}
\label{Subsec:Graph}

The thermodynamic function and the $N$-particle reduced density matrices can be calculated from the cluster functions $U^{(N_\uparrow, N_\downarrow)}$ through the primary $\zeta$-graphs and contracted $\zeta$-graphs introduced by Lee and Yang \cite{LY60-IV}. 
In this section, we show the main results which are used for later discussions. 
The details are shown in Appendix~\ref{App:LYCE}, where we use graph rules \cite{SKU12} different from those of Lee and Yang \cite{LY60-IV}. 
The relations between them are listed in Appendix~C of Ref.~\cite{SKU12}.

In terms of the primary or contracted $0$-graphs, 
we can write the grand partition function as
\begin{equation}
\begin{split}
 \ln \Xi_V 
& = \sum \left[ \text{all different primary $0$-graphs} \right] \\
& = \ln \Xi_{V, {\rm ideal}}
+ \mathcal{P},
\end{split}
\label{eq:XiGraph}
\end{equation}
where
\begin{equation}
 \mathcal{P}
 = \sum \left[ \text{all different contracted $0$-graphs} \right].
\end{equation}
Here, the definitions of the primary and contracted $0$-graphs are given in Appendix~\ref{App:LYCE}. 
The important point is that each primary or contracted $0$-graphs is computed from the cluster functions $U^{(N_\uparrow, N_\downarrow)}$.
By using Eq.~(\ref{eq:pressure}) (or $\beta \Delta p = \lim_{V \to \infty} \mathcal{P}/V$) and Eq.~(\ref{eq:pressure_Cluster_Expansion}),
Eq.~(\ref{eq:XiGraph}) gives the cluster expansion of the thermodynamic function.

\begin{figure}
\begin{picture}(250,97)
\put(5,-3){\includegraphics[width=235pt]{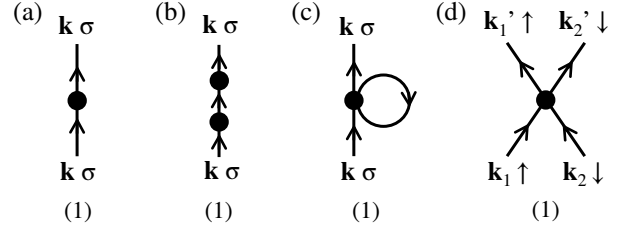}}
\end{picture}
\caption{
Primary $1$-graphs and $2$-graphs up to the second-order terms in fugacity $z$.
The graph in (a) contributes to the first-order term in the primary $1$-graphs.
The graphs in (b) and (c) contribute to the second-order term in the primary $1$-graphs.
The graph in (d) contribute to the second-order term in the primary $2$-graphs.
The symmetry number is shown under each graph.
}
\label{fig:Primary_12Graphs}
\end{figure}

The examples of the Lee-Yang primary $0$-graphs and $\mathcal{P}$
are illustrated in Fig.~\ref{fig:Primary_Graphs} and Fig.~\ref{fig:Xi}-(a), respectively.
Here, solid lines represent the primary graphs, and dotted lines represent the contracted graphs and describe the effect of the Fermi-Dirac statistics.
To be concrete, we consider the geometric series 
\begin{equation}
  \eta_0 (\bold{k}) := \sum_{n=0}^{\infty} \left[ - z \, e^{-\beta \hbar^2\bold{k}^2/(2m)} \right]^n
   = 1 - n_F( \bold{k} ),
\label{eq:PhysMeanOfEta0}
\end{equation}
where
$ n_F( \bold{k} ) :=   [ 1+ z^{-1} e^{\beta \hbar^2\bold{k}^2/(2m)} ]^{-1}$
is the Fermi distribution function.
The effect of Fermi-Dirac statistics emerges through this geometric series.
This sum is illustrated in Fig.~\ref{fig:Xi}-(b). 

Similarly, in terms of the primary $1$-graphs and $2$-graphs, 
we can write the single-particle and two-particle reduced density matrices \cite{dD62, SKU12}.
In particular, we have
\begin{equation}
\begin{split}
& \langle \hat{n}_{\bold{k} \sigma} \rangle 
 = \sum \left[ \text{all different primary $1$-graphs} \right], \\
\end{split}
\label{eq:nkGraph}
\end{equation}
and
\begin{equation}
\begin{split}
& \langle \hat{n}_{\bold{k} \uparrow} \hat{n}_{\bold{k}' \downarrow} \rangle 
- \langle \hat{n}_{\bold{k} \uparrow} \rangle \langle \hat{n}_{\bold{k}' \downarrow} \rangle \\
& = \sum \left[ \text{all different primary $2$-graphs} \right],
\end{split}
\label{eq:nknkGraph}
\end{equation}
where $\hat{n}_{\bold{k} \sigma}$ is the occupation number for particles with momentum $\bold{k}$ and spin $\sigma$ and $\langle \quad \rangle$ refers to the statistical average over a grand canonical ensemble.
The examples of the Lee-Yang primary $1$-graphs and $2$-graphs are illustrated in Fig.~\ref{fig:Primary_12Graphs} up to the second-order terms in fugacity $z$.



\section{Two-particle cluster function and second-order cluster expansion}
\label{sec:Crossover}

\subsection{Model potential}

As stated in the introduction, we consider a dilute gas of two-component fermions with zero-range attractive $s$-wave interaction.
There is no interaction between particles of the same spin due to the Pauli exclusion principle
and an interaction potential between particles of opposite spins $v \left(\bold{r}_i-\bold{r}_j \right)$ is characterized by the $s$-wave scattering length $a$.
Our Hamiltonian is given by
\begin{equation}
\begin{split}
 H^{(N_\uparrow, N_\downarrow)} = 
& - \sum_{i=1}^{N}  \frac{\hbar^2}{2m} \nabla_{i}^2 
 + \sum_{i=1}^{N_\uparrow} \sum_{j=N_\uparrow +1}^{N}
 \!\! v \left(\bold{r}_i-\bold{r}_j \right).
\label{eq:Hamiltonian}
\end{split}
\end{equation}


There are several forms of the potential for the zero-range interaction which reproduce the low-energy scattering properties.
One is the regularized $s$-wave pseudopotential \cite{Huang} 
\begin{equation}
v \left(\bold{r} \right) 
= \frac{4 \pi \hbar^2 a}{m} \delta \left(\bold{r} \right) \frac{\partial}{\partial r}( r \, \cdot  ),
\label{eq:def:pseudopotential}
\end{equation}
where $r = \left|\bold{r} \right|$, with $\bold{r} := \bold{r}_2 - \bold{r}_1$ being the relative coordinate.
To obtain the momentum representation of the pseudopotential,
we introduce Tan's $\Lambda$ function \cite{Tan08-a}
\begin{equation}
\delta \left(\bold{r} \right) \frac{\partial}{\partial r}( r e^{-i \bold{k} \cdot \bold{r}} )
\equiv \delta \left(\bold{r} \right) \Lambda \left(\bold{k} \right),
\label{eq:Lambda_fn}
\end{equation}
which has the following two properties:
\begin{align}
& \Lambda \left(\bold{k} \right) = 1 \qquad \text{for} \,\,\, \left|\bold{k} \right| < \infty ; \\
%
& \frac{1}{V} \sum_{\bold{k}}  \frac{\Lambda \left(\bold{k} \right)}{k^2}   = 0.
\label{eq:Lambda-fn-prop}
\end{align}
The explicit form of the $\Lambda$ function is given by Valiente \cite{V12} as
$ \Lambda \left(\bold{k} \right) = 1- k^{-1} \delta \left( k^{-1} \right) $.
Then, we obtain
\begin{equation}
\begin{split}
& \langle\bold{k}_1'; \bold{k}_2'| \,v\, |\bold{k}_1; \bold{k}_2\rangle \\
& := \frac{1}{V^2} \int_V \! d^3 \bold{R} \int_V \! d^3 \bold{r} \, e^{i (\bold{K}' \cdot \bold{R} + \bold{k}' \cdot \bold{r})} v \left(\bold{r} \right)  e^{-i (\bold{K} \cdot \bold{R} + \bold{k} \cdot \bold{r})} \\ 
& = \frac{1}{V} \, \delta_{\bold{K},\bold{K}'} \frac{4 \pi \hbar^2 a}{m} \Lambda \left(\bold{k} \right),
\label{eq:pseudopotential_momentum}
\end{split}
\end{equation}
where $\bold{K}=\bold{k}_{1}+\bold{k}_{2}$, $\bold{K}'=\bold{k}_{1}'+\bold{k}_{2}'$, 
$\bold{k}=(\bold{k}_{1}-\bold{k}_{2})/2$, and $\bold{k}'=(\bold{k}_{1}'-\bold{k}_{2}')/2$.
We note that the use of the $s$-wave pseudopotential is equivalent to imposing the Bethe-Peierls boundary condition \cite{Zwerger12, GPS08}:
\begin{equation}
\frac{1}{r \psi} \frac{d}{dr} (r \psi) \Big|_{r \to +0} = - \frac{1}{a}.
\label{eq:B-P}
\end{equation}

The Hamiltonian (\ref{eq:Hamiltonian}) with the $s$-wave pseudopotential (\ref{eq:def:pseudopotential}) supports continuous scattering states
\begin{equation}
\begin{split}
  \psi_{\rm sc}(r)=&\left\{ 2\pi^2[1+(k_{\rm sc}a)^2] \right\}^{-1/2} \\
& \times  \frac{1}{r}
    \bigl(  \sin (k_{\rm sc}r) -k_{\rm sc}a \cos (k_{\rm sc}r)  \bigr),
\label{eq:def:scat-w.f.}
\end{split}
\end{equation}
with the energy eigenvalue $E_{\rm sc} = \hbar^2 k_{\rm sc}^2/m$. 
Moreover, if $a>0$, the Hamiltonian also supports one bound state
\begin{equation}
\psi_{b}(r)= (2\pi a)^{-\frac{1}{2}} \, \frac{1}{r} \, e^{-r/a},
\label{eq:def:bound-w.f.}
\end{equation}
with the binding energy $E_b = - \hbar^2/(ma^2)$.

\subsection{Two-particle cluster function}

The exact two-particle cluster function $U^{(1,1)}$ can be obtained from the set of the energy eigenstates (\ref{eq:def:scat-w.f.}) and (\ref{eq:def:bound-w.f.}) and the corresponding eigenvalues.
Now, the two-particle Hamiltonian is
\begin{equation}
H^{(1,1)}= -\frac{\hbar^2}{2m} \nabla_1^2 -\frac{\hbar^2}{2m} \nabla_2^2 + v (\bold{r}) = -\frac{\hbar^2}{4m}\nabla_{\bold{R}}^2+H^{({\rm rel})},
\end{equation}
where
\begin{equation}
H^{({\rm rel})} = - \frac{\hbar^2}{m}\nabla_{\bold{r}}^2 + v  (\bold{r}).
\end{equation}
The two-particle cluster function for a finite volume $V$ is
\begin{equation}
\begin{split}
& \langle\bold{k}_1'; \bold{k}_2'|U^{(1,1)}|\bold{k}_1; \bold{k}_2\rangle  \\
& = \frac{8\pi^3}{V} \delta_{\bold{K},\bold{K}'} \,
     e^{- \beta \hbar^2 \bold{K}^2/(4m)} 
     \langle\bold{k}'| u^{({\rm rel})}  |\bold{k}\rangle,
\label{eq:U2_pseudopotential_finiteV}
\end{split}
\end{equation}
where
\begin{equation}
\langle\bold{k}'| u^{({\rm rel})}  |\bold{k}\rangle
:= \langle\bold{k}'| e^{-\beta H^{({\rm rel})}}  |\bold{k}\rangle -  \delta_{\bold{k},\bold{k}'} \, e^{-\beta \hbar^2 \bold{k}^2/m}.
\label{eq:Urel} 
\end{equation}
Here $\bold{K}:=\bold{k}_1+\bold{k}_2$ and $\bold{k}:=(\bold{k}_1-\bold{k}_2)/2$, and the Kronecker delta $\delta_{\bold{K},\bold{K}'}$ reflects the conservation of momentum.

The function $\langle\bold{k}'| u^{({\rm rel})}  |\bold{k}\rangle$ describes an effect of interaction (\ref{eq:def:pseudopotential}) and can be calculated from the eigenfunctions and eigenvalues of $H^{({\rm rel})}$ as shown in Appendix~\ref{app:U2_pseudo}. 
The result is 
\begin{equation}
  \langle\bold{k}'|u^{({\rm rel})}|\bold{k}\rangle  =
\begin{cases}
\displaystyle  \frac{\lambda^3}{2^{5/2}\pi^{7/2}} \, \frac{s(x', w) -  s(x, w)}{x'^2-x^2},
 & \text{for} \,\, x\not=x' ;
\\
\displaystyle  \frac{\lambda^3}{(2\pi)^{7/2}} \, \frac{1}{x} \frac{\partial}{\partial x} s(x, w),
 & \text{for} \,\, x=x' ,
\end{cases}
\label{eq:Urel_pseudo}
\end{equation}
where
\begin{equation}
\begin{split}
s(x, & w)
 =  \frac{1}{x^2+w^2} \\
& \times  \left(
     w \, e^{-x^2}
     -  \frac{2}{\sqrt{\pi}} \, x F \left(x \right)      
     -  w \, e^{w^2} \operatorname{erfc}  \left( -w \right)
    \right).
\label{eq:def:sx}
\end{split}
\end{equation}
Here we have introduced the dimensionless variables
$x:=\sqrt{\beta \hbar^2\bold{k}^2/m}= \lambda \left|\bold{k} \right|/\sqrt{2\pi}$ and $w:=\sqrt{\beta}\hbar/(\sqrt{m}a)=\lambda/(\sqrt{2\pi}a)$,
Dawson's integral 
\begin{equation}
F \left( x \right) =  e^{-x^2} \int_0^{x} dt \, e^{t^2},
\label{eq:def:DawsonF}
\end{equation}
 and the complementary error function
\begin{equation}
\operatorname{erfc} \left(x \right)
= \frac{2}{\sqrt{\pi}} \int_{x}^{\infty} \! dt \, e^{-t^2}.
\label{eq:def:erfc}
\end{equation}

The two-particle cluster function in the zero-range model was first obtained in Ref.~\cite{OU06} for $\lambda /a \geq 0$ and Ref.~\cite{In09} for $\lambda /a \! <0$.
However the expressions in Eqs.~(\ref{eq:U2_pseudopotential_finiteV})-(\ref{eq:def:sx}) hold for either sign of $\lambda /a$
and the two-particle cluster function is smoothly connected at the unitary limit $\lambda /a=0$.
%

\subsection{Second-order cluster integral}

The second cluster integral $\Delta b_2$ is calculated from the graph illustrated as Fig.~\ref{fig:Primary_Graphs}-(c).
The algebraic expression of Fig.~\ref{fig:Primary_Graphs}-(c) is given in Appendix~\ref{App:Alg-Exp}-1.
From Eq.~(\ref{App:eq:Db2}), we have
\begin{equation}
\frac{V}{\lambda^3} \Delta b_2
 = \sum_{\bold{k}_1, \bold{k}_2}
   \langle\bold{k}_1; \bold{k}_2|U^{(1,1)}|\bold{k}_1; \bold{k}_2\rangle .
\label{eq:b2-1}
\end{equation}
Using Eq.~(\ref{eq:U2_pseudopotential_finiteV}) with Eqs.~(\ref{eq:Urel_pseudo}) and (\ref{eq:def:sx}),
the right-hand side (RHS) of Eq.~(\ref{eq:b2-1}) is rewritten as
\begin{equation}
\begin{split}
&  \frac{8\pi^3}{V} \sum_{\bold{K}} e^{- \beta\hbar^2\bold{K}^2/(4m)} 
      \sum_{\bold{k}}  \langle\bold{k}| u^{({\rm rel})}  |\bold{k}\rangle \\
& = \frac{V}{\lambda^3} \sqrt{2} \, e^{w^2}  \operatorname{erfc} \left( - w \right).
\end{split}
\end{equation}
Thus, we obtain
\begin{equation}
\Delta b_2
  = \sqrt{2} \, e^{w^2}  \operatorname{erfc} \left( - w \right) . 
\label{eq:b2-BethUhl}
\end{equation}
This result is nothing but the Beth-Uhlenbeck formula \cite{HM04}.
From this result, it is confirmed that the cluster-expansion method can treat effects of quantum-mechanical scattering and bound states at least at the level of the second cluster integral.

The third cluster integral $\Delta b_3$ can be calculated from the graphs illustrated as Fig.~\ref{fig:Primary_Graphs}-(e) and (f).
The algebraic expressions of Fig.~\ref{fig:Primary_Graphs}-(e) and (f) are given in Appendix~\ref{App:Alg-Exp}-1.
The calculation of the graph in Fig.~\ref{fig:Primary_Graphs}-(e) can be done in a manner similar to that of $\Delta b_2$.
However, the calculation of the graph in Fig.~\ref{fig:Primary_Graphs}-(f) is not straightforward, because it includes the functions $U^{(1,2)}$ and $U^{(2,1)}$, which are obtained by a solution of a three-body problem.
Recently, the calculation of the term corresponding to Fig.~\ref{fig:Primary_Graphs}-(f) and hence of $\Delta b_3$ was carried out by Leyronas \cite{L11} for an arbitrary $\lambda /a$. 
Here we simply comment on the correspondence between the Lee-Yang primary $0$-graphs in Fig.~\ref{fig:Primary_Graphs} and the graphs used in Ref.~\cite{L11}.
Note that the Lee-Yang primary $0$-graphs represent the terms that appear in $\ln \Xi$,
but the graphs used in Ref.~\cite{L11} represent the terms that appear in the number density.
According to Eqs.~(\ref{eq:pressure}) and (\ref{eq:densitiy}),
the differential of the former is equivalent to the latter.
The Lee-Yang primary $0$-graphs in Fig.~\ref{fig:Primary_Graphs}-(a), (b), (c), and (f) correspond to the graphs in Fig.~1, Fig.~2-(a), Fig.~2-(b), and Fig.~5 in Ref.~\cite{L11}, respectively.
The Lee-Yang primary $0$-graphs in Fig.~\ref{fig:Primary_Graphs}-(e) correspond to the sum of the graphs in Fig.~3 and Fig.~4 in Ref.~\cite{L11}.

\subsection{Second-order contact coefficient}

Since we have obtained $\Delta b_2$ in Eq.~(\ref{eq:b2-BethUhl}), 
we can calculate the second-order contact coefficient $c_2$ by using Eq.~(\ref{eq:def:ContactCoeff}) as
\begin{equation}
c_2 = \frac{1}{\sqrt{2 \pi}} \frac{d}{dw} \Delta b_2
 = \frac{2}{\pi} + \frac{2}{\sqrt{\pi}} w \, e^{w^2}  \operatorname{erfc} \left( - w \right). 
\label{eq:c2}
\end{equation}
Here, we calculate $c_2$ by means of the thermodynamic quantity $b_2$.
Beside thermodynamic quantities, Tan's contact can be calculated by means of the average occupation number $\langle \hat{n}_{\bold{k} \sigma} \rangle$ in the momentum space at large momentum $1/k^4$ tail as in Eq.~(\ref{eq:def:contact}) and by means of the pair correlation at short distances as  \cite{Tan08-a}
\begin{equation}
\begin{split}
\left<  \hat{n}_\uparrow(\bold{r}) \hat{n}_\downarrow(\bold{0}) \right> 
&= \frac{C}{16\pi^2} \left(\frac{1}{r^2}- \frac{2}{ar}\right) + O\left( r^0 \right) .
\end{split}
\label{eq:Contact-by-shortdis}
\end{equation}
Here, $\left< \hat{n}_\uparrow(\bold{r}) \hat{n}_\downarrow(\bold{0}) \right>$ is the statistical average of the density-density correlation with definite positions $\bold{r}$ with spin $\uparrow$ and $\bold{0}$ with spin $\downarrow$ in an infinite volume over the grand canonical ensemble. 
We devote the rest of this subsection to rederive $c_2$ by means of these two method, and demonstrate that the cluster-expansion method can give the $1/k^4$ asymptotic behavior of 
$\langle \hat{n}_{\bold{k} \sigma} \rangle$ at large momentum and the $1/r^2$ asymptotic behavior of $\left<  \hat{n}_\uparrow(\bold{r}) \hat{n}_\downarrow(\bold{0}) \right> $ at short distances, at least up to the second order in fugacity.




We calculate $\langle \hat{n}_{\bold{k} \uparrow} \rangle$ at large momentum up to the second-order terms in fugacity.
According to Eq.~(\ref{eq:nkGraph}), we have
\begin{equation}
\begin{split}
\langle \hat{n}_{\bold{k} \uparrow} \rangle = 
& z \, e^{- \beta \hbar^2 \bold{k}^2/(2m)} - z^2 e^{- \beta \hbar^2 \bold{k}^2/m} \\
&+ z^2 \sum_{\bold{q}} 
     \langle \bold{k} ; \bold{q} | U^{(1,1)} | \bold{k} ; \bold{q} \rangle
 + O\left( z^3 \right),
\end{split}
\end{equation}
in which each term corresponds to the primary $1$-graphs in Fig.~\ref{fig:Primary_12Graphs}-(a), (b), and (c).
For $\lambda k \gg 1$, i.e., $x \gg 1$, the asymptotic behavior of Dawson's integral is
$F \left( x \right) = 1/(2x) + O\left( 1/x^3 \right)$.
Thus, we have
\begin{equation}
s(x) =  \frac{-1}{x^2} 
  \left[ \frac{1}{\sqrt{\pi}}  + w \, e^{w^2} \operatorname{erfc}  \left( -w \right)  \right] 
      + O\left( \frac{1}{x^4} \right) .
\label{eq:sx-asymptotic}
\end{equation}
By substituting Eq.~(\ref{eq:sx-asymptotic}) into Eq.~(\ref{eq:U2_pseudopotential_finiteV}) and using Eq.~(\ref{eq:Urel_pseudo}),
we have
\begin{equation}
\begin{split}
& \sum_{\bold{q}} 
     \langle \bold{k} ; \bold{q} | U^{(1,1)} | \bold{k} ; \bold{q} \rangle \\
& =   \frac{2^6}{(\lambda k)^4}   \left[
       1 + \sqrt{\pi} \, w \, e^{w^2} \operatorname{erfc}  \left( -w \right)  \right]
      + O\left( \frac{1}{(\lambda k)^6} \right).
\end{split}
\end{equation}
Therefore, we obtain
\begin{equation}
\begin{split}
C & \equiv \lim_{\lambda k\to\infty} k^4 \langle \hat{n}_{\bold{k} \uparrow} \rangle \\
  & = \frac{2^6 z^2}{\lambda^4} \left[
      1 + \sqrt{\pi} \, w \, e^{w^2} \operatorname{erfc}  \left( -w \right)  \right] 
       + O\left( z^3 \right).
\label{eq:contact_by_momentum}
\end{split}
\end{equation}
Comparing Eq.~(\ref{eq:contact_by_momentum}) with Eq.~(\ref{eq:CE_of_contact}),
we obtain $c_2$ which agrees with Eq.~(\ref{eq:c2}).

We calculate $\left< \hat{n}_\uparrow(\bold{r}) \hat{n}_\downarrow(\bold{0}) \right>$ at short distance up to the second-order terms in fugacity.
According to the Fourier transform of Eq.~(\ref{eq:nknkGraph}), we have
\begin{equation}
\begin{split}
&\left<  \hat{n}_\uparrow(\bold{r}) \hat{n}_\downarrow(\bold{0}) \right> 
- \left<  \hat{n}_\uparrow(\bold{r}) \right>\left< \hat{n}_\downarrow(\bold{0}) \right>  
\\
&= z^2 \langle \bold{r} ; \bold{0} | U_{\infty}^{(1,1)} | \bold{r} ; \bold{0} \rangle
 + O\left( z^3 \right),
\end{split}
\end{equation}
in which the first term on the right-hand side corresponds to the Fourier transform of the primary $2$-graphs in Fig.~\ref{fig:Primary_12Graphs}-(d). [See Eq.~(17) in Ref.~\cite{SKU12}.]
By using $\left<  \hat{n}_\uparrow(\bold{r}) \right> = \left< \hat{n}_\downarrow(\bold{0}) \right>
= z / \lambda^3 +  O( z^2 )$,
which is obtained by the Fourier transform of 
$\langle \hat{n}_{\bold{k} \uparrow} \rangle = \langle \hat{n}_{\bold{k} \downarrow} \rangle = z \, e^{- \beta \hbar^2 \bold{k}^2/(2m)} + O( z^2 )$,
and Eq.~(\ref{eq:U2-infry}) given in Appendix~\ref{app:U2_pseudo},
we have
\begin{equation}
\left<  \hat{n}_\uparrow(\bold{r}) \hat{n}_\downarrow(\bold{0}) \right> 
= \frac{z^2}{\lambda^6} 
 + \frac{2^{3/2}z^2}{\lambda^3}   \langle\bold{r} | u^{({\rm rel})}_{\infty}  |\bold{r}\rangle
  + O\left( z^3 \right).
\end{equation}
Here, from Eq.~(\ref{eq:pseudo_real-sp.2}) given in Appendix~\ref{app:U2_pseudo}, we have
\begin{equation}
\begin{split}
  \langle\bold{r}| u^{({\rm rel})}_{\infty}| \bold{r}\rangle 
  = &  \frac{1}{2^{3/2}\lambda} 
    \left( \frac{1}{\pi} + \frac{1}{\sqrt{\pi}} w \, e^{w^2}\operatorname{erfc} \left( -w \right) \right)  \\
& \times  \left(\frac{1}{r^2}- \frac{2}{ar}\right) + O\left( r^0 \right).
\end{split}
\end{equation}
Therefore, we obtain
\begin{equation}
\begin{split}
&\left<  \hat{n}_\uparrow(\bold{r}) \hat{n}_\downarrow(\bold{0}) \right> \\
&= \frac{z^2}{\lambda^4} 
    \left( \frac{1}{\pi}+ \frac{1}{\sqrt{\pi}} w \, e^{w^2}\operatorname{erfc} \left( -w \right) \right)\left(\frac{1}{r^2}- \frac{2}{ar}\right) \\
& \quad  + O\left( r^0 \right) + O\left( z^3 \right).
\end{split}
\label{eq:TwoBodyCorr-short}
\end{equation}
Comparing Eqs.~(\ref{eq:Contact-by-shortdis}) and (\ref{eq:TwoBodyCorr-short}) with Eq.~(\ref{eq:CE_of_contact}),
we obtain $c_2$ which agrees with Eq.~(\ref{eq:c2}).

\section{Phase transition temperature in the BCS and BEC limits}
\label{sec:PT}

\subsection{Identification of the phase transition point in terms of cluster expansion}



A phase transition manifests itself as the appearance of a singularity in the thermodynamic function \cite{Huang}.
Here, the singularity is defined by the disappearance of holomorphy (or analyticity)\footnote{
A complex-valued function is said to be holomorphic on an open set $\Omega$ in the complex plane, 
if its Taylor expansion around any point in $\Omega$ has a nonzero radius of convergence. 
If $\Omega$ is not an open set, we interpret that holomorphy holds in an appropriate open set containing $\Omega$.}.
In this subsection, we describe the statistical theory of phase transitions in the context of our model.

The cluster expansion of the equation of state (in parametric form) is written as
Eqs.~(\ref{eq:pressure_Cluster_Expansion}) and (\ref{eq:density_Cluster_Expansion}).
We fix the $s$-wave scattering length $a$ and the temperature $T$.
Then, the cluster integrals $\{ b_n \}_{n=1,2,\dots}$ are fixed,
because they depend only on $\lambda/a$. 
To identify the phase transition point, we consider a singularity of the RHS of Eq.~(\ref{eq:pressure_Cluster_Expansion}) near the origin and along the positive real axis in the complex $z$-plane: $0 \leq z < \infty$.
It is reasonable to assume that our system has a point $z_c$ on the positive real axis so that the RHS of Eq.~(\ref{eq:pressure_Cluster_Expansion}) is holomorphic at $0 \leq z < z_c$ and has singularity 
 at $z=z_c$.
It indicates that the system is in the normal gas phase on $0 \leq z < z_c$, 
and at $z=z_c$ there is a phase transition, which is often identified with the superfluid phase transition \cite{
GPS08, Zwerger12}.
%
%
%
%
%
%
Substituting $z_c$ to Eq.~(\ref{eq:density_Cluster_Expansion}),
we obtain the value $(\rho \lambda^3 )_c$ at the phase transition point.
Then, we obtain the transition temperature $T_c$, using 
\begin{equation}
\frac{T_c}{T_F} = \frac{4}{3^{2/3} \pi^{1/3} [(\rho \lambda^3 )_c]^{2/3}}.
\label{eq:TcTf_by_rho}
\end{equation}
Similarly, we obtain $(k_F a)^{-1}$ at the corresponding point by using the relation $k_F = (3 \pi^2 \rho )^{1/3}$.

For our system, 
\begin{equation}
\sum_{n=1}^\infty b_n z^n = - 2 \operatorname{Li}_{\frac{5}{2}} (-z) + \sum_{n=2}^\infty \Delta b_n z^n ,
\end{equation}
and
there is no singularity in $-2 \operatorname{Li}_{5/2} (-z)$ near the origin and along the positive real axis in the complex $z$-plane.
Thus, to examine a phase transition, we examine a singularity of $\sum_{n=2}^\infty \Delta b_n z^n$ on the positive real axis of $z$.
Using $\sum_{n=2}^\infty \Delta b_n z^n = \lambda^3 \lim_{V \to \infty} \mathcal{P} /V$,
we can evaluate the critical point through the Lee-Yang contracted graphs.

We comment on the correspondence between the above procedure and the procedure of the approximate BCS-BEC crossover theory by Nozi\`eres and Schmitt-Rink (NSR) \cite{NSR85, MRE93}.
In the theory by NSR, the Thouless criterion and the number equation are solved simultaneously to calculate the transition temperature $T_c$.
In our theory, the Thouless criterion corresponds to the determination of the singularity of Eq.~(\ref{eq:pressure_Cluster_Expansion}),
and the number equation is replaced by Eq.~(\ref{eq:density_Cluster_Expansion}).

\subsection{Pairing approximation}

The discussion in the previous subsection is general. 
Here we develop an approximate theory that satisfies the following two requirements:
(i) in the weak-coupling limit, the transition temperature $T_c$ is consistent with the BCS theory;
(ii) in the strong-coupling limit, $T_c$ reduces to that of BEC of non-interacting tightly bound dimers.
The transition temperature is determined by different physical mechanisms in the weak- and strong-coupling limits:
(i) in the weak-coupling limit,  $T_c$ is determined by the Cooper instability of the Fermi sphere;
(ii) in the strong-coupling limit, $T_c$ is determined by the onset of BEC of dimers in the zero center-of-mass state.
Thus, we must take into account the quantum-exchange effect of the Fermi-Dirac statistics of particles and the quantum-exchange effect of the Bose statistics of pairs.
In this paper, we do not consider the Gor'kov--Melik-Barkhudarov correction \cite{GM61}
, which is important in the weak-coupling regime ($(k_F a)^{-1} \lesssim -1$), and the scattering between dimers, which is important in the strong-coupling regime ($(k_F a)^{-1} \gtrsim 1$) \cite{KPS01, PSS04}.

\begin{figure}
\begin{picture}(250,125)
\put(10,-5){\includegraphics[width=220pt]{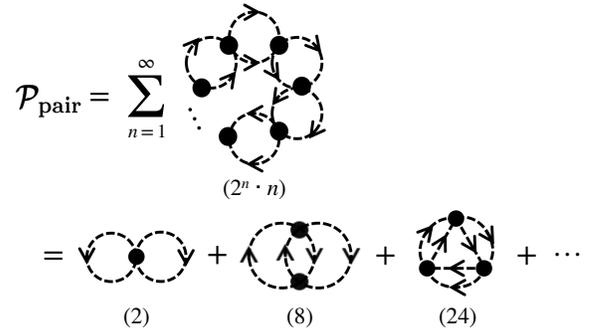}}
\end{picture}
\caption{
Expression of $\mathcal{P}_{\rm pair}$ 
as the sum over different ladder-type contracted $0$-graphs.
The corresponding symmetry numbers are shown under the graphs.}
\label{fig:XiLad}
\end{figure}

To meet the above requirements,
we consider a set of contracted $0$-graphs $\mathcal{P}_{{\rm pair}}$ as shown in Fig.~\ref{fig:XiLad},
and approximate the grand partition function as
\begin{equation}
\ln \Xi_V \simeq 
- 2 \frac{V}{\lambda^3} \operatorname{Li}_{\frac{5}{2}} (-z) 
+ \mathcal{P}_{{\rm pair}}.
\label{GPF_Ppair}
\end{equation}
The algebraic expression of $\mathcal{P}_{{\rm pair}}$ is given in Appendix~\ref{App:Alg-Exp}-2.
From Eq.~(\ref{App:eq:XiLad-Irr}), we have
\begin{equation}
\begin{split}
&  \mathcal{P}_{{\rm pair}}
 =  \sum_{n=1}^\infty \frac{z^{2n}}{n}   \sum_{\bold{k}_1,\dots,\bold{k}_{2n}} 
   \prod_{i=1}^n  \left( 1 - n_F(\bold{k}_{2i-1}) \right) \\
& \quad \times \left( 1 - n_F(\bold{k}_{2i}) \right)  
 \langle\bold{k}_{2i+1}; \bold{k}_{2i+2}|U^{(1,1)}|\bold{k}_{2i-1}; \bold{k}_{2i}\rangle, 
\label{eq:XiLad-Irr}
\end{split}
\end{equation}
where $\bold{k}_{2n+1} := \bold{k}_{1}$ and $\bold{k}_{2n+2} := \bold{k}_{2}$.
We call the above approximation (\ref{GPF_Ppair}) the ``pairing approximation''.
The physical meaning of the pairing approximation is quite simple.
The sum $\mathcal{P}_{{\rm pair}}$ includes the effect of the Fermi sphere $( 1 - n_F(\bold{k}) )$, that of two-particle scattering (and pairing in the case of a positive scattering length) with opposite spins $\langle\bold{k}_1'; \bold{k}_2' |U^{(1,1)}|\bold{k}_1; \bold{k}_2 \rangle$,
and that of the Bose statistics of pairs of particles with opposite spins (for details, see Example~4 in page~9 in Ref.~\cite{SKU12}).

However, it is unclear whether or not the correct transition temperature can be derived from the pairing approximation, even in the weak- and strong-coupling limits.
The present study is the first attempt to investigate the BCS-BEC crossover based on the Lee-Yang cluster-expansion method.
In the following subsections C and D, we derive the transition temperature under pairing approximation in the weak- and strong-coupling limits.
This shows that the pairing approximation satisfies the requirements in the first paragraph of this subsection.

Let us rewrite Eq.~(\ref{eq:XiLad-Irr}) for use in subsections C and D.
Substituting Eq.~(\ref{eq:U2_pseudopotential_finiteV}) into Eq.~(\ref{eq:XiLad-Irr}),
we obtain
\begin{equation}
\begin{split}
 \mathcal{P}_{{\rm pair}}
 = & \sum_{\bold{K}}
 \sum_{n=1}^\infty \frac{1}{n} \left( \frac{8 \pi^3 z^2}{V} \right)^n
     e^{- n \beta \hbar^2\bold{K}^2/(4m)}   \\
& \times     \sum_{\bold{p}_1,\dots,\bold{p}_{n}}    
  \prod_{i=1}^n
 \left( 1 - n_F\left( {\textstyle \frac{1}{2}} \bold{K} + \bold{p}_i \right) \right) \\
& \times \left( 1 - n_F\left( {\textstyle \frac{1}{2}} \bold{K} - \bold{p}_i \right) \right) 
  \langle\bold{p}_{i+1} |u^{({\rm rel})}|\bold{p}_{i} \rangle,
\label{eq:XiLad-Irr-final1}
\end{split}
\end{equation}
where $\bold{p}_{i} := (\bold{k}_{2i} - \bold{k}_{2i-1})/2$ for $i=1,\dots, n$ and $\bold{p}_{n+1} := \bold{p}_{1}$.
%
%
Using 
\begin{equation}
\begin{split}
& z^2 e^{\beta \hbar^2\bold{K}^2/(4m)}  
 \left( 1 - n_F\left( \textstyle\frac{1}{2} \bold{K} + \bold{p}_i \right) \right) 
 \left( 1 - n_F\left( \textstyle\frac{1}{2} \bold{K} - \bold{p}_i \right) \right) \\
& =  e^{\beta \hbar^2\bold{p}_i^2/m} \,
 n_F \left( \textstyle\frac{1}{2} \bold{K} + \bold{p}_i \right)
 n_F \left( \textstyle\frac{1}{2} \bold{K} - \bold{p}_i \right) ,
\end{split}
\end{equation}
we obtain
\begin{equation}
\begin{split}
 \mathcal{P}_{{\rm pair}}
 = & \sum_{\bold{K}}
 \sum_{n=1}^\infty \frac{1}{n} \left( \frac{8 \pi^3}{V} \right)^n \!\!
    \sum_{\bold{p}_1,\dots,\bold{p}_{n}}     
  \prod_{i=1}^n
 n_F \left( {\textstyle \frac{1}{2}} \bold{K} + \bold{p}_i \right) \\
&\times n_F \left( {\textstyle \frac{1}{2}} \bold{K} - \bold{p}_i \right) 
   e^{\beta \hbar^2\bold{p}_i^2/m}
 \langle\bold{p}_{i+1} |u^{({\rm rel})}|\bold{p}_{i} \rangle.
\label{eq:XiLad-Irr-final2}
\end{split}
\end{equation}



\subsection{Weak-coupling limit (BCS limit)}

In this subsection, we shall evaluate the transition temperature under pairing approximation (\ref{GPF_Ppair}) in the weak-coupling limit ($(k_F a)^{-1} \ll -1$).
In this limit, we take into account the two-particle cluster function $U^{(1,1)}$ 
up to the leading order of the $s$-wave scattering length $a$ as
\begin{equation}
\begin{split}
& \langle\bold{k}_{1}'; \bold{k}_{2}'|U^{(1,1)}|\bold{k}_{1}; \bold{k}_{2}\rangle  
 = \int_0^\beta \!\! d\tau \,
 e^{-(\beta-\tau)(\epsilon_{\bold{k}_{1}'}+\epsilon_{\bold{k}_{2}'})}  \\
& \quad \times \langle\bold{k}_{1}'; \bold{k}_{2}'| \,v\, |\bold{k}_{1}; \bold{k}_{2}\rangle  
 e^{-\tau (\epsilon_{\bold{k}_{1}}+\epsilon_{\bold{k}_{2}})} 
  + O\left( \left( a/\lambda \right) ^2\right),
\label{eq:1st-order_of_U2}
\end{split}
\end{equation}
where $\epsilon_{\bold{k}_i} = \hbar^2\bold{k}^2_i/(2m)$ and $\epsilon_{\bold{k}_i'} = \hbar^2\bold{k}_i^{\prime 2}/(2m)$ for $i=1,2$.
%
%
Substituting Eq.~(\ref{eq:pseudopotential_momentum}) into the RHS of Eq.~(\ref{eq:1st-order_of_U2}),
and comparing the result with Eq.~(\ref{eq:U2_pseudopotential_finiteV}), we obtain
\begin{equation}
 \langle\bold{k}'| u_{\rm 1st}^{({\rm rel})}  |\bold{k}\rangle
  = 
  \frac{\hbar^2 a}{2\pi^2 m}  \Lambda \left(\bold{k} \right)
\int_0^\beta \!\! d\tau \,
 e^{-(\beta-\tau)\hbar^2\bold{k}'^2/m}
 e^{-\tau \hbar^2\bold{k}^2/m}.
\label{eq:Urel_pseudopotential_1st}
\end{equation}
Substituting Eq.~(\ref{eq:Urel_pseudopotential_1st}) into Eq.~(\ref{eq:XiLad-Irr-final2}),
we obtain
\begin{equation}
\begin{split}
 \mathcal{P}_{{\rm pair}}^{{\rm 1st}}
 = & \sum_{\bold{K}}
 \sum_{n=1}^\infty \frac{1}{n}  \left( \frac{4\pi \hbar^2 a}{Vm} \right)^n 
    \sum_{\bold{p}_1,\dots,\bold{p}_{n}}     
 \prod_{i=1}^n
  n_F \left(  {\textstyle \frac{1}{2}} \bold{K} + \bold{p}_i \right)\\
& \times n_F \left(  {\textstyle \frac{1}{2}} \bold{K} - \bold{p}_i \right)
   \Lambda \left(\bold{p}_i \right) 
  \int_0^\beta \!\! d\tau_i \,
 e^{\tau_i \hbar^2(\bold{p}_{i+1}^2 - \bold{p}_i^2 )/m}.
\label{eq:XiLad-Irr-1st}
\end{split}
\end{equation}
Here the first-order approximation of $\mathcal{P}_{{\rm pair}}$ is denoted by $ \mathcal{P}_{{\rm pair}}^{{\rm 1st}} $.
Using the property of Tan's $\Lambda$ function (\ref{eq:Lambda-fn-prop}),
it is shown in Appendix~\ref{app:Weak coupling limit} that
\begin{equation}
 \mathcal{P}_{{\rm pair}}^{{\rm 1st}}
 =  \sum_{l \in \mathbb{Z}} \sum_{\bold{K}}
 \sum_{n=1}^\infty \frac{1}{n}    
  \left[ Q(\bold{K}, \Omega_l) \right]^n  ,
\label{BCS-limit_InfSum}
\end{equation}
where
\begin{equation}
\begin{split}
 Q(\bold{K}, \Omega_l)
 =  
  \frac{4\pi \hbar^2 a}{Vm}
  \sum_{\bold{p}}
  \left[
  \frac{1-n_F(\bold{k}_{1})-n_F(\bold{k}_{2})}{i\Omega_l -
 (\epsilon_{\bold{k}_{1}} + \epsilon_{\bold{k}_{2}} -2\mu)}
  + \frac{m}{\bold{p}^2}   \right] ,
\label{eq:Q_weak-limit}
\end{split}
\end{equation}
$\bold{k}_{1}= (1/2)\bold{K} +\bold{p}$, and $\bold{k}_{2}= (1/2)\bold{K} -\bold{p}$.
Here, $\Omega_l = 2\pi l /\beta$ is the bosonic Matsubara frequency and the summation $\sum_{l \in \mathbb{Z}}$ extends over all integers $l \in \{ 0, \pm 1, \pm 2, \dots \}$.

Since $\left| Q(\bold{K}, \Omega_l) \right| < Q(0, 0)$ for all $\bold{K} \not= 0$ and $\Omega_l \not= 0$,
the convergence of Eq.~(\ref{BCS-limit_InfSum}) is determined by the condition $Q(0, 0) < 1$.
By using Eq.~(\ref{eq:Q_weak-limit}) and $1-2 n_F(\bold{p})=\tanh [\beta(\epsilon_\bold{p} -\mu)/2]$,
we obtain
\begin{equation}
\begin{split}
Q(0, 0)  =
 - \frac{4\pi \hbar^2 a}{Vm}
  \sum_{\bold{p}}
  \left\{ 
  \frac{\tanh [\beta(\epsilon_\bold{p} -\mu)/2]}
  {2(\epsilon_\bold{p} -\mu)}
  - \frac{1}{2\epsilon_\bold{p}}   \right\}.
\end{split}
\end{equation}
Therefore, the convergence condition of Eq.~(\ref{BCS-limit_InfSum}) is written as $\beta < \beta_c$, where
\begin{equation}
\begin{split}
 - \frac{4\pi \hbar^2 a}{Vm}
  \sum_{\bold{p}}
  \left\{ 
  \frac{\tanh [\beta_c(\epsilon_\bold{p} -\mu)/2]}
  {2(\epsilon_\bold{p} -\mu)}
  - \frac{1}{2\epsilon_\bold{p}}   \right\} = 1.
\label{eq:Thouless_criterion}
\end{split}
\end{equation}
Here, $\beta_c$ is the inverse of the transition temperature.
In the weak-coupling limit, the chemical potential $\mu$ is equal to the Fermi energy of free fermions $\hbar^2 k_F^2/(2m)$ \cite{NSR85}.
Equation (\ref{eq:Thouless_criterion}) is equivalent to the Thouless criterion \cite{T60},
and gives the transition temperature of the BCS theory \cite{GM61, MRE93} as 
\begin{equation}
\frac{T_c}{T_F} \simeq 0.61 \exp \left( - \frac{\pi}{2 k_F |a|} \right).
\end{equation}

Equation (\ref{BCS-limit_InfSum}) shows a close connection between the present theory and the theory by NSR \cite{NSR85, MRE93}.
If we extrapolate the present theory to the strong-coupling regime, 
our theory reduces to that by NSR.
If the sum in Eq.~(\ref{BCS-limit_InfSum}) is convergent, i.e. $\beta < \beta_c$, we obtain
\begin{equation}
\begin{split}
& \mathcal{P}_{{\rm pair}}^{{\rm 1st}}
 =  - \sum_{l \in \mathbb{Z}} \sum_{\bold{K}}
 \ln  \left[ 1 - Q(\bold{K}, \Omega_l) \right]  \\
& = - \sum_{l \in \mathbb{Z}} \sum_{\bold{K}} 
 \ln \biggl\{    \frac{1}{V}
  \sum_{\bold{p}} 
 \left[
   \frac{1-n_F(\bold{k}_{1})-n_F(\bold{k}_{2})}{i\Omega_l -
 (\epsilon_{\bold{k}_{1}} + \epsilon_{\bold{k}_{2}} -2\mu)}
  + \frac{m}{\hbar^2\bold{p}^2} 
 \right] \\
& \qquad - \frac{m}{4\pi \hbar^2 a}\biggr\} 
  + \mathrm{constant}.
\label{BCS-limit_EoS}
\end{split}
\end{equation}
By using Eq.~(\ref{GPF_Ppair}) altogether with Eq.~(\ref{BCS-limit_EoS}), 
we obtain
$\ln \Xi_V \simeq 
\ln \Xi_{{\rm ideal}, V} 
+ \mathcal{P}_{{\rm pair}}^{{\rm 1st}}$.
Then, the number equation is given by
\begin{equation}
\rho \simeq 
\rho_{\rm ideal} 
+ z \frac{\partial}{\partial z} \lim_{V \to \infty} \frac{1}{V} \mathcal{P}_{{\rm pair}}^{{\rm 1st}}.
\label{eq:NumberEq}
\end{equation}
Equation (\ref{eq:NumberEq}) coincides with the number equation of  the NSR theory  \cite{NSR85, MRE93}.

It might appear that the present theory gives a BCS-BEC crossover theory based on the Lee-Yang cluster-expansion method,
because we have reproduced the Thouless criterion (\ref{eq:Thouless_criterion}) and the number equation (\ref{BCS-limit_EoS}),
which together produce the BCS-BEC crossover theory by NSR.
Then, following NSR \cite{NSR85, MRE93},
we obtain $T_c/T_F \simeq 0.22$ in the unitary limit $(k_F a)^{-1}=0$ and obtain the transition temperature of dimers in the strong-coupling limit.
However, the assumption in Eq.~(\ref{eq:1st-order_of_U2}) is valid only in the weak-coupling regime.
Therefore, we conclude that this result is not the derivation of the BCS-BEC crossover theory but that of the BCS theory only.
In the strong-coupling limit, we should treat two-particle cluster function $U^{(1,1)}$ nonperturbatively.
The next subsection is devoted to demonstrate the right way to obtain the transition temperature of dimers based on the Lee-Yang cluster-expansion method.

One may suspect that our theory is just a rewriting of the theory by NSR.
However, it is non-trivial to derive the Thouless criterion from the Lee-Yang cluster-expansion method in the weak-coupling limit.
The reason is the following.
The thermal Green function for a free fermion, which is used to derive the Thouless criterion, involves the information on the Fermi sphere.
However, in the formulation of a cluster expansion, the $n$-th cluster integral $b_n$ has only information of $n$ particles, and has no information of the Fermi sphere.
In the above discussion, we have overcome this difficulty by considering the pairing approximation and involving the Fermi-sphere effect of the surrounding fermions by using 
the contracted graph which is constructed from an infinite series of the primary graphs as shown in Fig.~\ref{fig:Xi}-(b).
While the above derivation invokes Tan's $\Lambda$ function (\ref{eq:Lambda_fn}), Eq.~(\ref{BCS-limit_EoS}) can also be obtained by using the delta-function-type contact interaction and the standard regularization
(see, Appendix~\ref{app:Weak coupling limit}-2).



\subsection{Strong coupling limit (BEC limit)}

In this subsection, we shall evaluate the transition temperature under pairing approximation (\ref{GPF_Ppair}) in the strong-coupling limit ($(k_F a)^{-1} \gg 1$)\footnote{
The following derivation of BEC of dimers was first obtained in Ref.~\cite{OU06}.
Here we rederive it to demonstrate the difference between our theory and the theory by NSR.}.
In this limit, we cannot treat two-particle cluster function $U^{(1,1)}$ perturbatively.
We first rewrite Eq.~(\ref{eq:Urel_pseudo}) as
\begin{equation}
\begin{split}
  \langle\bold{k}'|u^{({\rm rel})}|\bold{k}\rangle 
&  = 
\langle\bold{k}'|u^{({\rm rel, b})}|\bold{k}\rangle  \\
& \, +
\frac{\lambda^3}{2^{5/2}\pi^{7/2}} \, \frac{s_{{\rm sc}}(x',w) -  s_{{\rm sc}}(x,w)}{x'^2-x^2}  
,
\end{split}
\end{equation}
where
\begin{equation}
\begin{split}
& s_{\rm sc}(x,w)
 =  \frac{1}{x^2+w^2} \\
& \quad \times   \left(
    w e^{-x^2}
     -  \frac{2}{\sqrt{\pi}} \, x F \left(x \right)
     +  \left| w \right| e^{w^2} \operatorname{erfc}  \left( \left| w \right|  \right)
    \right),
\end{split}
\end{equation}
and
\begin{equation}
\langle\bold{k}'|u^{({\rm rel, b})}|\bold{k}\rangle 
= \theta(a) \cdot e^{\beta\hbar^2/(m a^2)} \psi_{b}(\bold{k}') \psi_{b}^*(\bold{k}) .
\label{eq:def:urel_b}
\end{equation}
Here
\begin{equation}
\theta \left( x \right) =  
\begin{cases}
1,  & \quad{\mathrm{for}} \,\,\, x > 0 , \\
0,  & \quad{\mathrm{for}} \,\,\,  x \leq 0
\end{cases}
\end{equation}
is the Heaviside step function
and
\begin{equation}
\psi_{b}(\bold{k}) = \frac{a^{3/2}}{\pi}\frac{1}{1+ (ka)^2}
\end{equation}
is the normalized relative wave function of the bound state, which is the Fourier transform of Eq.~(\ref{eq:def:bound-w.f.}).
Keeping the leading order term, we have
\begin{equation}
\begin{split}
\langle\bold{k}'|u^{({\rm rel})}|\bold{k}\rangle 
&  = 
\langle\bold{k}'|u^{({\rm rel, b})}|\bold{k}\rangle + O(a/\lambda)  \\
&   = e^{\beta\hbar^2/(m a^2)} 
\psi_{b}(\bold{k}') \psi_{b}(\bold{k}) + O(a/\lambda).
\label{eq:Urel_Strng-cpling}
\end{split}
\end{equation}
Substituting Eq.~(\ref{eq:Urel_Strng-cpling}) and $ 1- n_F(\bold{k}) \simeq 1$ into Eq.~(\ref{eq:XiLad-Irr-final1}),
we obtain
\begin{equation}
\begin{split}
 \mathcal{P}_{{\rm pair}} 
& = \sum_{\bold{K}}
 \sum_{n=1}^\infty \frac{1}{n}  \biggl[ z^2 e^{\beta\hbar^2/(m a^2)} e^{- \beta \hbar^2\bold{K}^2/(4m)}    \\
& \qquad \times \frac{8 \pi^3}{V}  \sum_{\bold{k}} \left| \psi_{b}(\bold{k}) \right|^2  \biggr]^n \\
& = 2 \sqrt{2} \frac{V}{\lambda^3} \operatorname{Li}_{\frac{5}{2}} \left(  z^2 e^{\beta\hbar^2/(m a^2)} \right) .
\label{BEC-limit_InfSum}
\end{split}
\end{equation}

The condition for convergence of Eq.~(\ref{BEC-limit_InfSum}) is $z^2 e^{\beta\hbar^2/(m a^2)} < 1$
which may also be rewritten as
\begin{equation}
2\mu <  2\mu_c \equiv - \frac{\hbar^2}{ma^2} = E_b  .
\end{equation}
The number equation (\ref{eq:density_Cluster_Expansion}) is
\begin{equation}
\begin{split}
\rho
 & \simeq \rho_{\rm ideal} 
+ z \frac{\partial}{\partial z} \lim_{V \to \infty} \frac{1}{V} \mathcal{P}_{{\rm pair}} \\
 & = - \frac{2}{\lambda^3} \operatorname{Li}_{\frac{3}{2}} \left(  -z \right) 
 +   \frac{4\sqrt{2}}{\lambda^3} \operatorname{Li}_{\frac{3}{2}} \left(  z^2 e^{\beta\hbar^2/(m a^2)} \right) .
\end{split}
\end{equation}
Here we neglect the free-particle part $\rho_{\rm ideal}$, because $\rho_{\rm ideal} \ll \rho - \rho_{\rm ideal}$ in the strong-coupling limit.
Therefore, at the transition temperature $T_c$,
$ \rho  \simeq 
 4\sqrt{2} \, \lambda_c^{-3} \operatorname{Li}_{3/2} \left( 1 \right)$,
where $\lambda_c$ is the corresponding thermal de Broglie length at $T_c$.
Using Eq.~(\ref{eq:TcTf_by_rho}), we obtain
\begin{equation}
\begin{split}
\frac{T_c}{T_F}
\simeq  \left[ \frac{2}{ 9 \pi \left( \zeta \left(  \frac{3}{2} \right) \right)^2} \right]^{1/3} 
\simeq 0.2180,
\label{eq:TcOfDimerBEC}
\end{split}
\end{equation}
where $\zeta (x)$ is the Riemann zeta function and $\zeta(3/2)=\operatorname{Li}_{3/2}(1) = 2.612\ldots$.
This result is identical to the transition temperature for non-interacting diatomic molecules $T_c = (\pi \hbar^2 /m)[ \rho / (2 \, \zeta \left( 3/2 \right))]^{2/3}$.


Note that we have obtained the transition temperature of dimers (\ref{eq:TcOfDimerBEC}) in a manner fundamentally different from the NSR theory \cite{NSR85, MRE93}.
In Sec.~\ref{sec:PT}-C, we have derived the Thouless criterion (\ref{eq:Thouless_criterion}) and the number equation (\ref{BCS-limit_EoS}).
Then, following the procedure by NSR with the extrapolation,
in the strong-coupling limit we obtain Eq.~(\ref{eq:TcOfDimerBEC}).
In other words, in the strong-coupling limit, we have obtained the same result (\ref{eq:TcOfDimerBEC}) based on two different approximations.
However, as discussed in the last of Sec.~\ref{sec:PT}-C, Eqs.~(\ref{eq:Thouless_criterion}) and (\ref{BCS-limit_EoS}) are derived with the weak-coupling approximation, and it seems rather fortuitous that this approach also reproduces the correct result of Eq.~(\ref{eq:TcOfDimerBEC}).

\section{Conclusion and future prospects}

In this paper, we have proposed a new approach to the BCS-BEC crossover based on the cluster-expansion method of Lee and Yang.
We have evaluated the transition temperature of this system, by analyzing an emergence of the singularity of an infinite series of cluster functions.
We have shown that an infinite sum of cluster functions $\mathcal{P}_{{\rm pair}}$ has the following three properties: 
(i) in the weak-coupling limit, it gives the Thouless criterion and hence the transition temperature of BCS theory, and the number equation of the BCS-BEC crossover theory by NSR;
(ii) in the strong-coupling limit, it reproduces the thermodynamic function of non-interacting tightly bound dimers; 
(iii) in the high-temperature limit, it gives the exact second virial coefficient, which is valid also in the unitary regime.
All of these suggest that our theory provides a good starting point for a new BCS-BEC crossover theory and this is the first work to demonstrate how to derive these three limits from a unified point of view.

\begin{figure}
\begin{picture}(250,120)
\put(10,-5){\includegraphics[width=230pt]{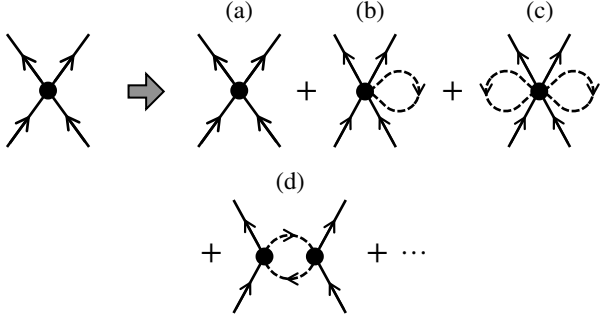}}
\end{picture}
\caption{
Graphical representation of medium effects on a binary collision.
The left-hand side of the rightwards thick arrow represent a bare binary collision,
and the right-hand side represent a binary collision with medium effects.
In this figure, we display all graphs up to the fourth-order in fugacity $z$.
The orders of the fugacity are $z^2$ for (a), $z^3$ for (b), and $z^4$ for (c) and (d).
%
}
\label{fig:future}
\end{figure}

Finally, we discuss two future prospects about the present approach to the BCS-BEC crossover.
(i) One is to compute the transition temperature $T_c/T_F$ for an arbitrary $s$-wave scattering length $(k_F a)^{-1}$ under the pairing approximation (\ref{GPF_Ppair}).
We have not done that, because we have not yet understood how to carry out the infinite summation $\mathcal{P}_{{\rm pair}}$ in Eq.~(\ref{eq:XiLad-Irr}), which is necessary to obtain the number equation.
(ii) The other is to take into account the medium effects for the transition temperature due to the Gor'kov--Melik-Barkhudarov (GMB) correction \cite{GM61} in the weak-coupling regime ($(k_F a)^{-1} \lesssim -1$) and the scattering between dimers in the strong-coupling regime ($(k_F a)^{-1} \gtrsim 1$) \cite{KPS01, PSS04}.
Both of them are caused by the medium particles surrounding a quantum-condensed pair boson (a Cooper pair or a dimer boson), e.g., the GMB correction is caused by the screening of the interaction strength by the presence of virtual particle-hole excitations.
In the sum $\mathcal{P}_{{\rm pair}}$ in Eq.~(\ref{eq:XiLad-Irr}),
the $2$-vertex corresponds to a bare binary collision without medium effects.
Then, in order to take into account the medium effects, we have to include graphs which correspond to a binary collision with medium effects.
In Fig.~\ref{fig:future}, we display graphical representation of medium effects on a binary collision\footnote{The infinite sum of the right-hand side in Fig.~\ref{fig:future} is called a simple $2$-diagram in Ref.~\cite{dD62} and is called an irreducible contracted 2-graph in Ref.~\cite{SKU12}.}.
The replacement as Fig.~\ref{fig:future} in the sum $\mathcal{P}_{{\rm pair}}$ may lead to incorporating the medium effects.

\begin{acknowledgments}
We thank K.~Inokuchi, and N.~Kawakami for useful discussions.
M.~U. acknowledges the financial support by a Grant-in-Aid (KAKENHI~22340114 and 22103005) and the Photon Frontier Network Program, from MEXT of Japan.
N.~S. was supported by a Grant-in-Aid for JSPS Fellows (Grant No.~250588) and Y.~N. was supported by JSPS KAKENHI Grant Number 25887020.
\end{acknowledgments}

\appendix

\section{Lee-Yang cluster-expansion method}
\label{App:LYCE}

In this appendix, we review the cluster-expansion method of Lee and Yang \cite{LY58-I, LY60-IV, SKU12} in a system of two-component fermions described in Sec.~\ref{sec:LYCE}-A.
In this method, the grand partition function and the $N$-particle reduced density matrices are expressed in terms of the primary $\zeta$-graphs or the contracted $\zeta$-graphs to be defined later.
Each primary or contracted $\zeta$-graph is computed from the cluster functions $U^{(N_\uparrow, N_\downarrow)}$ for the same system obeying Boltzmann statistics, which is defined in Sec.~\ref{sec:LYCE}-B.
In particular, in this appendix we define the primary $\zeta$-graphs and the contracted $\zeta$-graphs, and show how to express the grand partition function and the $N$-particle reduced density matrices in terms of the primary or contracted $\zeta$-graphs.
As remarked in Ref. \cite{SKU12}, we use the rules different from those of Lee and Yang \cite{LY60-IV}.
The relations between Lee-Yang and our notation are listed in Appendix~C of Ref. \cite{SKU12}.

\subsection{Definition of antisymmetric combination $\Upsilon_{\rm A}$}

In the computation of the Lee-Yang cluster expansion method for fermions, only the antisymmetric combination of $U^{(l_\uparrow, l)}$ appears.
Then, we define the matrix elements of $\Upsilon_{\rm A}^{(l_\uparrow, l_\downarrow)}$ by
\begin{equation}
\begin{split}
& \langle 1',\dots , l_\uparrow'; (l_\uparrow+1)',\dots , l'
| \Upsilon_{\rm A}^{(l_\uparrow, l_\downarrow)}  | 
  1,\dots , l_\uparrow; l_\uparrow +1,\dots , l \rangle \\
& :=  \sum_{P \in S_{l_\uparrow}} \sum_{Q \in S_{l_\downarrow}} (-1)^P (-1)^Q 
   \bigl\langle 1',\dots , l_\uparrow'; (l_\uparrow+1)',\dots , l \big|  \\
& \qquad U^{(l_\uparrow, l_\downarrow)} 
  \big|\, P(1),\dots , P(l_\uparrow); Q(l_\uparrow +1),\dots , Q(l)  \, \bigr\rangle ,
\end{split}
\label{eq:def:UpsA}
\end{equation}
where $l=l_\uparrow + l_\downarrow$.
Here, $P$ and $Q$ denote permutations among up-spin and down-spin particles, respectively,
and $(-1)^P$ and $(-1)^Q$ take on $1$ or $-1$ for even or odd permutations.

It is useful to define a function $\Upsilon_{\rm A}^{(l)}$ related to
$\Upsilon_{\rm A}^{(l_\uparrow, l_\downarrow)}$ as follows.
First, we define
\begin{equation}
\begin{split} 
& \langle q_1', \dots ,q_l' |  \Upsilon_{\rm A}^{(l)}
  |  q_1, \dots ,q_l \rangle  
 := \\
 & \langle 1',\dots , l_\uparrow'; (l_\uparrow+1)',\dots , l' | 
 \Upsilon_{\rm A}^{(l_\uparrow, l_\downarrow)}  | 
  1,\dots , l_\uparrow; l_\uparrow +1,\dots , l \rangle ,
\end{split}
\end{equation}
where
$q_1 := (\bold{k}_1, \uparrow)$, \dots ,
$q_{l_\uparrow} := (\bold{k}_{l_\uparrow}, \uparrow)$,
$q_{l_\uparrow +1} := (\bold{k}_{l_\uparrow +1}, \downarrow)$,
 \dots ,
$q_{l} := (\bold{k}_{l}, \downarrow)$,
and
$q_1' := (\bold{k}_1', \uparrow)$, \dots ,
$q_{l_\uparrow}' := (\bold{k}_{l_\uparrow}', \uparrow)$,
$q_{l_\uparrow +1}' := (\bold{k}_{l_\uparrow +1}', \downarrow)$,
 \dots ,
$q_{l}' := (\bold{k}_{l}', \downarrow)$.
Here $l:=l_\uparrow + l_\downarrow$.
Then,
\begin{equation}
\begin{split}
& 
\langle Q(q_1'), \dots ,Q(q_l') | \Upsilon_{\rm A}^{(l)}
  |  P(q_1), \dots ,P(q_l) \rangle  \\
& := (-1)^P (-1)^Q 
 \langle q_1', \dots ,q_l' |  \Upsilon_{\rm A}^{(l)}
  |  q_1, \dots ,q_l \rangle .
\end{split}
\end{equation}

We give a few examples.\\

\noindent
\textit{Example 1} (one particle):
\begin{gather}
    \langle \bold{k}' , \uparrow | \Upsilon_{\rm A}^{(1)} | \bold{k} , \uparrow \rangle
    = \langle \bold{k}' | \Upsilon_{\rm A}^{(1,0)} | \bold{k} \rangle
    = \langle \bold{k}' | U^{(1,0)} | \bold{k} \rangle.  \\
    \langle \bold{k}' , \downarrow | \Upsilon_{\rm A}^{(1)} | \bold{k} , \downarrow \rangle
    = \langle \bold{k}' | \Upsilon_{\rm A}^{(0,1)} | \bold{k} \rangle.
    = \langle \bold{k}' | U^{(0,1)} | \bold{k} \rangle.  \\
    \langle \bold{k}' , \uparrow | \Upsilon_{\rm A}^{(1)} | \bold{k} , \downarrow \rangle
    = \langle \bold{k}' , \downarrow | \Upsilon_{\rm A}^{(1)} | \bold{k} , \uparrow \rangle
    = 0  .
\end{gather}

\noindent
\textit{Example 2} (two particles with opposite spins):
\begin{equation}
\begin{split}
  & \quad   \langle \bold{k}'_1, \uparrow; \bold{k}'_2, \downarrow | \Upsilon_{\rm A}^{(2)} | \bold{k}_1, \uparrow; \bold{k}_2, \downarrow \rangle \\
  &   =  -
  \langle \bold{k}'_2, \downarrow; \bold{k}'_1, \uparrow | \Upsilon_{\rm A}^{(2)} | \bold{k}_1, \uparrow; \bold{k}_2, \downarrow \rangle \\
  &   =  -
  \langle \bold{k}'_1, \uparrow; \bold{k}'_2, \downarrow | \Upsilon_{\rm A}^{(2)} | \bold{k}_2, \downarrow; \bold{k}_1, \uparrow \rangle \\
  &  =  \langle \bold{k}'_2, \downarrow; \bold{k}'_1, \uparrow | \Upsilon_{\rm A}^{(2)} | \bold{k}_2, \downarrow; \bold{k}_1, \uparrow \rangle \\
  &   =  \langle \bold{k}'_1; \bold{k}'_2 | \Upsilon_{\rm A}^{(1,1)} | \bold{k}_1; \bold{k}_2 \rangle  
      =  \langle \bold{k}'_1; \bold{k}'_2 | U^{(1,1)} | \bold{k}_1; \bold{k}_2 \rangle .
\end{split}
\end{equation}

\noindent
\textit{Example 3} (two particles with the same spin):
\begin{equation}
\begin{split}
  &    \langle \bold{k}'_1, \sigma; \bold{k}'_2, \sigma | \Upsilon_{\rm A}^{(2)} | \bold{k}_1, \sigma; \bold{k}_2, \sigma \rangle \\
  &   =  - 
  \langle \bold{k}'_2, \sigma; \bold{k}'_1, \sigma | \Upsilon_{\rm A}^{(2)} | \bold{k}_1, \sigma; \bold{k}_2, \sigma \rangle \\
  &   =  - 
  \langle \bold{k}'_1, \sigma; \bold{k}'_2, \sigma | \Upsilon_{\rm A}^{(2)} | \bold{k}_2, \sigma; \bold{k}_1, \sigma \rangle \\
  &  =  \langle \bold{k}'_2, \sigma; \bold{k}'_1, \sigma | \Upsilon_{\rm A}^{(2)} | \bold{k}_2, \sigma; \bold{k}_1, \sigma \rangle \\
  & =
   \begin{cases}
     \langle \bold{k}'_1, \bold{k}'_2 | \Upsilon_{\rm A}^{(2,0)} | \bold{k}_1, \bold{k}_2 \rangle ,
      & \quad {\rm for} \,\,\, \sigma = \uparrow ; \\
     \langle \bold{k}'_1, \bold{k}'_2 | \Upsilon_{\rm A}^{(0,2)} | \bold{k}_1, \bold{k}_2 \rangle ,
      & \quad {\rm for} \,\,\, \sigma = \downarrow .
   \end{cases}
\end{split}
\end{equation}
Here,
\begin{equation}
\begin{split}
  \langle \bold{k}'_1, \bold{k}'_2 | \Upsilon_{\rm A}^{(2,0)} | \bold{k}_1, \bold{k}_2 \rangle
= & \langle \bold{k}'_1, \bold{k}'_2 | U^{(2,0)} | \bold{k}_1, \bold{k}_2 \rangle  \\
  & - \langle \bold{k}'_1, \bold{k}'_2 | U^{(2,0)} | \bold{k}_2, \bold{k}_1 \rangle ,
\end{split}
\end{equation}
and
\begin{equation}
\begin{split}
  \langle \bold{k}'_1, \bold{k}'_2 | \Upsilon_{\rm A}^{(0,2)} | \bold{k}_1, \bold{k}_2 \rangle
= & \langle \bold{k}'_1, \bold{k}'_2 | U^{(0,2)} | \bold{k}_1, \bold{k}_2 \rangle  \\
  & - \langle \bold{k}'_1, \bold{k}'_2 | U^{(0,2)} | \bold{k}_2, \bold{k}_1 \rangle .
\end{split}
\end{equation}


\subsection{Thermodynamic function and reduced density matrices in terms of Lee-Yang primary $\zeta$-graphs}
\label{subsec:PriG}

The cluster expansion of the thermodynamic function and the $N$-particle reduced density matrices can be expressed in terms of the sum over connected products of  $\Upsilon_{\rm A}^{(l)}$ functions.
The exact character of this sum is most simply described in terms of primary $\zeta$-graphs or contracted $\zeta$-graphs introduced by Lee and Yang \cite{LY60-IV}.
A primary $\zeta$-graph is defined as follows:\\

\textit{Definition.}---
A primary $\zeta$-graph ($\zeta = 0,1,2, \dots$) is a graphical structure which consists of a collection of vertices connected by directed lines, with $\zeta$ external incoming lines and $\zeta$ external outgoing lines.
Here, a line that has vertices at both ends is called an internal line; otherwise, it is called an external line.
All external lines are considered distinguishable.
Each vertex, called the $l$-vertex ($l=1,2,\dots$), 
connects $l$ incoming lines and $l$ outgoing lines.
A primary $\zeta$-graph must include at least one vertex and one line, and all parts must be connected (i.e., there must be a path from any one vertex to any other vertex).
Two primary graphs are different if their topological structures are different.

The examples of the Lee-Yang primary $\zeta$-graphs are illustrated in Fig.~\ref{fig:Primary_Graphs} and \ref{fig:Primary_12Graphs}. 
To each of these graphs we assign a term which is determined by the following procedures:
\begin{enumerate}
\item[(i)] Associate with each internal line a different integer $i \; (i=1,\dots,N)$ 
and the corresponding coordinate and spin $q_i = (\bold{k}_i,\sigma_i)$.
Associate with each external line some prescribed coordinate and spin.
\item[(ii)] Assign to each $l$-vertex, a factor 
\begin{equation}
\begin{picture}(38,20)
\put(-2,17){\small $B_1$}
\put(5,-12){\line(5,4){30}}
\put(-2,-21){\small $A_1$}
\put(12,-6){\vector(4,3){1.5}}
\put(29,7){\vector(4,3){1.5}}
\put(34,17){\small $B_l$}
\put(35,-12){\line(-5,4){30}}
\put(34,-21){\small $A_l$}
\put(28,-6){\vector(-4,3){1.5}}
\put(11,7){\vector(-4,3){1.5}}
\put(20,0){\line(-2,-5){4.8}}
\put(17,-8){\vector(1,2){1.5}}
\put(10,-21){\small $A_2$}
\put(22,-21){\small $\dots$}
\put(20,0){\line(-2,5){4.8}}
\put(17,7){\vector(-1,2){1.5}}
\put(10,17){\small $B_2$}
\put(22,17){\small $\dots$}
\put(20,0){\circle*{5}}
\end{picture}
 = z^l 
\langle B_1, \dots, B_l |  \Upsilon_{\rm{A}}^{(l)} | A_1, \dots, A_l \rangle,
\end{equation}\\ 
where $A_i$ and $B_i$ represent the coordinates and spins associated with the incoming and outgoing $i$th lines $(i=1,\dots,l)$, respectively.
The number of up-spins associated with its incoming lines is the same as that of outgoing lines,
and it is denoted as  $l_\uparrow$.
The same is true for down-spins $l_\downarrow$, with $l=l_\uparrow + l_\downarrow$.

\item[(iii)] Assign a factor $1/S$ to the entire graph,
where $S$ is the symmetry number and is defined as follows:

Consider all $N!$ permutations of the positions of $N$ integers associated with the internal lines.
The total number of permutations that leave the graph topologically
unchanged gives the symmetry number of the graph.
The symmetry numbers are listed under each graph 
in Figs.~\ref{fig:Primary_Graphs}, \ref{fig:Xi}, \ref{fig:Primary_12Graphs}, and \ref{fig:XiLad}.
\item[(iv)] Assign a factor $-1$ to the entire graph,
if the permutation 
\begin{equation}
A_1 \rightarrow B_1,\; A_2\rightarrow B_2,\; \dots ,\; A_i \rightarrow B_i,\;\dots
\end{equation}
from all the initial coordinates into all the final coordinates of all the vertex function 
$\Upsilon^{(l)}_{\rm{A}}$ taken together is odd. 
\end{enumerate}

The term that corresponds to each graph is given by
\begin{equation}
\begin{split}
  \sum_{q_1,\dots, q_l} 
  [ {\rm \, product\;of\;all\;factors\;in\; (ii)-(iv) \,} ].
 \label{eq:Pri-Term}
\end{split}
\end{equation}


In terms of these primary $0$-graphs, 
we can write the grand partition function \cite{LY60-IV} as
\begin{equation}
\begin{split}
& \ln \Xi_V 
 = \sum \left[ \text{all different primary $0$-graphs} \right], \\
\end{split}
\label{eq:XiPri}
\end{equation}
to which each graph contributes a term given by Eq.~(\ref{eq:Pri-Term}).
A concrete calculation is given in Sec. III-C.
By using Eqs.~(\ref{eq:pressure}) and (\ref{eq:pressure_Cluster_Expansion}),
Eq.~(\ref{eq:XiPri}) gives the cluster expansion of the thermodynamic function.

Similarly, in terms of these primary $1$-graphs and $2$-graphs, 
we can write the single-particle and two-particle reduced density matrices, respectively.
For example, we have\footnote{For the derivation of Eq.~(\ref{eq:1part-Pri}), see Eq.~(IV.84) in Ref.~\cite{LY60-IV}. 
For  the derivation of Eq.~(\ref{eq:2part-Pri}), see Eqs.~(41) and (45) in Ref.~\cite{SKU12}.}
\begin{equation}
\begin{split}
& \langle \hat{n}_{\bold{k} \sigma} \rangle 
 = \sum \left[ \text{all different primary $1$-graphs} \right], \\
\end{split}
\label{eq:1part-Pri}
\end{equation}
and
\begin{equation}
\begin{split}
& \langle \hat{n}_{\bold{k} \uparrow} \hat{n}_{\bold{k}' \downarrow} \rangle 
- \langle \hat{n}_{\bold{k} \uparrow} \rangle \langle \hat{n}_{\bold{k}' \downarrow} \rangle \\
& = \sum \left[ \text{all different primary $2$-graphs} \right].
\end{split}
\label{eq:2part-Pri}
\end{equation}

\subsection{Thermodynamic function in terms of Lee-Yang contracted  $0$-graphs}

It is convenient to introduce a contracted $\zeta$-graph. 
A contracted $\zeta$-graph has the same topological structure 
as a primary $\zeta$-graph except that a contracted graph does not have any $1$-vertex. 
To each contracted graph, 
we assign a term which is determined by the same procedures
(i)-(iv) and the following additional rule:
\begin{enumerate}
\item[(v)] Assign a factor 
\begin{equation}
 \eta_{0}( \bold{k}_i ) := 1 - n_F( \bold{k}_i )
 =   \left[ 1+ z\,e^{-\beta \hbar^2\bold{k}_i^2/(2m)} \right]^{-1}
\label{eq:def:eta_0}
\end{equation}
to the $i$th internal line.
\end{enumerate}

In terms of the contracted $0$-graphs, Eq.~(\ref{eq:XiPri}) is expressed as
\begin{equation}
 \ln \Xi_V = \ln \Xi_{V, {\rm ideal}}
+ \mathcal{P},
\label{eq:logX-contracted}
\end{equation}
where
\begin{equation}
 \mathcal{P}
 = \sum \left[ \text{all different contracted $0$-graphs} \right].
\label{eq:def:MathcalP}
\end{equation}
By using $\beta \Delta p = \lim_{V \to \infty} \mathcal{P}/V$ and Eq.~(\ref{eq:Cluster_Expansion_Delta_p}),
Eq.~(\ref{eq:def:MathcalP}) gives the cluster expansion of the thermodynamic function.

Here $\mathcal{P}$ is illustrated in Fig.~\ref{fig:Xi}-(a).
The number under each graph in Fig.~\ref{fig:Xi}-(a) shows the symmetry number of the corresponding contracted $0$-graph.
The algebraic expression of the sum of the contracted graphs is
\begin{equation}
\begin{split}
  \mathcal{P}
& =  \frac{z^2}{2}
   \sum_{q_1,q_2}
     \eta_0 (\bold{k}_1) \eta_0 (\bold{k}_{2}) 
 \langle q_1,q_2 | \Upsilon_{\rm{A}}^{(2)} | q_1, q_2 \rangle  \\
&  + \frac{z^3}{6}
 \! \sum_{q_1,q_2,q_3} \!\!
     \eta_0 (\bold{k}_1) \eta_0 (\bold{k}_{2})  \eta_0 (\bold{k}_{3}) 
 \langle q_1,q_2,q_3 | \Upsilon_{\rm{A}}^{(3)} | q_1, q_2, q_3 \rangle  \\
&  +  \frac{z^{4}}{2}
 \! \sum_{q_1,\dots,q_4}\!
    \eta_0 (\bold{k}_1)  \eta_0 (\bold{k}_2) \eta_0 (\bold{k}_3) \eta_0 (\bold{k}_4) \\
& \qquad \times 
  \langle q_1,q_2 | \Upsilon_{\rm{A}}^{(2)} | q_1,q_3 \rangle  
     \langle q_3,q_4 | \Upsilon_{\rm{A}}^{(2)} | q_2,q_4\rangle \\
%
&  +  \frac{z^{4}}{8}
 \!\sum_{q_1,\dots ,q_4}\! 
    \eta_0 (\bold{k}_1)  \eta_0 (\bold{k}_2) \eta_0 (\bold{k}_3) \eta_0 (\bold{k}_4) \\
& \qquad \times 
  \langle q_1,q_2 | \Upsilon_{\rm{A}}^{(2)} | q_3,q_4 \rangle  
     \langle q_3,q_4 | \Upsilon_{\rm{A}}^{(2)} | q_1,q_2 \rangle   \\
&  + \cdots, 
\end{split}
\end{equation}
where each term in the sum corresponds to the contracted $0$-graph in the same order as in Fig.~\ref{fig:Xi}-(a).
Here, $ \eta_{0}( \bold{k} )$ describes the effect of the Fermi-Dirac statistics described below Eq.~(\ref{eq:PhysMeanOfEta0}).

\section{Algebraic expressions of the Lee-Yang graphs in Figs.~\ref{fig:Primary_Graphs} and \ref{fig:XiLad}}
\label{App:Alg-Exp}

In this appendix, we give the algebraic expressions of the Lee-Yang graphs that appear in Figs.~\ref{fig:Primary_Graphs} and \ref{fig:XiLad} in the system described in Sec.~\ref{sec:Crossover}, whose Hamiltonian is given in Eq.~(\ref{eq:Hamiltonian}).
The number under each term in Figs.~\ref{fig:Primary_Graphs} and \ref{fig:XiLad} is the symmetry number of the corresponding primary or contracted $0$-graph.
Since the particles with the same spin do not interact, $U^{(2,0)}=U^{(0,2)}=0$ and $U^{(3,0)}=U^{(0,3)}=0$.

\subsection{Algebraic expressions of Lee-Yang graphs in Fig.~\ref{fig:Primary_Graphs}}

The first cluster integral $b_1$ is calculated from the graph illustrated in Fig.~\ref{fig:Primary_Graphs}-(a).
The algebraic expression is
\begin{equation}
\begin{split}
& \frac{V}{\lambda^3} b_1
 = \sum_{q}
   \langle q |\Upsilon_{\rm A}^{(1)}| q \rangle \\
& = \sum_{\bold{k}} \left(
   \langle\bold{k} |U^{(1,0)}|\bold{k} \rangle 
 +  \langle\bold{k} |U^{(0,1)}|\bold{k} \rangle \right) 
  = 2 \frac{V}{\lambda^3}.
\end{split}
\end{equation}

The second cluster integral $\Delta b_2$ is calculated from the graph illustrated in Fig.~\ref{fig:Primary_Graphs}-(c).
The algebraic expression is
\begin{equation}
\begin{split}
\frac{V}{\lambda^3} \Delta b_2
& = \frac{1}{2} \sum_{q_1, q_2}
   \langle q_1, q_2|\Upsilon_{\rm A}^{(2)}| q_1, q_2 \rangle \\
& = \sum_{\bold{k}_1, \bold{k}_2}
   \langle\bold{k}_1; \bold{k}_2|U^{(1,1)}|\bold{k}_1; \bold{k}_2\rangle .
\label{App:eq:Db2}
\end{split}
\end{equation}
Here, we use $U^{(2,0)}=U^{(0,2)}=0$.

The third cluster integral $\Delta b_3$ is calculated from the graphs illustrated as Fig.~\ref{fig:Primary_Graphs}-(e) and (f).
The algebraic expression is
\begin{equation}
\Delta b_3 = \Delta b_3^{\text{(e)}} + \Delta b_3^{\text{(f)}},
\end{equation}
where
\begin{equation}
\begin{split}
& \frac{V}{\lambda^3} \Delta b_3^{\text{(e)}}
 = \sum_{q_1, q_2, q_3}
   \langle q_1, q_3|\Upsilon_{\rm A}^{(2)}| q_1, q_2 \rangle
   \langle q_2 |\Upsilon_{\rm A}^{(1)}| q_3 \rangle \\
& = \sum_{\bold{k}_1, \bold{k}_2, \bold{k}_3} \Bigl(
   \langle\bold{k}_1 ; \bold{k}_3 |U^{(1,1)}|\bold{k}_1; \bold{k}_2 \rangle
   \langle\bold{k}_2 |U^{(0,1)}|\bold{k}_3\rangle \\
   & \qquad \qquad +
     \langle\bold{k}_3 ; \bold{k}_1 |U^{(1,1)}|\bold{k}_2; \bold{k}_1 \rangle
   \langle\bold{k}_2 |U^{(1,0)}|\bold{k}_3\rangle
    \Bigr),
\end{split}
\end{equation}
and
\begin{equation}
\begin{split}
\frac{V}{\lambda^3} \Delta b_3^{\text{(f)}}
& = \frac{1}{6} \sum_{q_1, q_2, q_3}
   \langle q_1, q_2, q_3|\Upsilon_{\rm A}^{(3)}| q_1, q_2, q_3 \rangle \\
& = \frac{1}{2} \sum_{\bold{k}_1, \bold{k}_2, \bold{k}_3} \Bigl(
   \langle\bold{k}_1, \bold{k}_2; \bold{k}_3 |U^{(2,1)}|\bold{k}_1, \bold{k}_2; \bold{k}_3\rangle \\
   & \qquad \qquad +
     \langle\bold{k}_1; \bold{k}_2, \bold{k}_3 |U^{(1,2)}|\bold{k}_1; \bold{k}_2, \bold{k}_3\rangle \Bigr).
\end{split}
\end{equation}
Here, we use $U^{(2,0)}=U^{(0,2)}=0$ and $U^{(3,0)}=U^{(0,3)}=0$.

\subsection{Algebraic expressions of $\mathcal{P}_{\rm pair}$ in Fig.~\ref{fig:XiLad}}

We consider a set of contracted $0$-graphs $\mathcal{P}_{{\rm pair}}$ as shown in Fig.~\ref{fig:XiLad}.
The algebraic expression of $\mathcal{P}_{{\rm pair}}$ is written as 
\begin{equation}
\begin{split}
  \mathcal{P}_{{\rm pair}}
 = 
&  \sum_{n=1}^{\infty} \frac{z^{2n}}{n \cdot 2^n}  \sum_{q_1,\dots,q_{2n}}
  \prod_{i=1}^n  \eta_0 (\bold{k}_{2i-1}) \eta_0 (\bold{k}_{2i})  \\
&  \times  \langle q_{2i+1}, q_{2i+2}| \Upsilon_{\rm{A}}^{(2)} | q_{2i-1}, q_{2i}\rangle .
\label{eq:def:P_pair}
\end{split}
\end{equation}
By using $U^{(2,0)}=U^{(0,2)}=0$, the RHS of Eq.~(\ref{eq:def:P_pair}) is
\begin{equation}
\begin{split}
& \sum_{n=1}^\infty \frac{z^{2n}}{n}   \sum_{\bold{k}_1,\dots,\bold{k}_{2n}}
  \prod_{i=1}^n  \eta_0 (\bold{k}_{2i-1}) \eta_0 (\bold{k}_{2i})  \\
& \times \langle\bold{k}_{2i+1}; \bold{k}_{2i+2}|U^{(1,1)}|\bold{k}_{2i-1}; \bold{k}_{2i}\rangle,
\label{App:eq:XiLad-Irr}
\end{split}
\end{equation}
where $\bold{k}_{2n+1} := \bold{k}_{1}$ and $\bold{k}_{2n+2} := \bold{k}_{2}$.

\section{Derivation of the two-particle cluster function for the $s$-wave pseudopotential}
\label{app:U2_pseudo}

\subsection{Two-particle cluster function for the $s$-wave pseudopotential}

To calculate the two-particle cluster function $U^{(1,1)}$ in the $s$-wave approximation, the general formula discussed in Ref.~\cite{LY58-I} is applied.
The formula deals with the case of a central potential and an infinite volume $V=\infty$.
The relationship between the cases of finite $V$ and infinite $V$
is discussed in Appendix E of Ref. \cite{LY60-IV}.
In this appendix,
we show the subscripts $V$ and $\infty$ to distinguish the cases of finite $V$ and infinite $V$, respectively.

The general formula in Ref.~\cite{LY58-I} requires a complete set of energy eigenvalues and eigenfunctions,
which are calculated by using the pseudopotential (\ref{eq:def:pseudopotential}).
The two-particle Hamiltonian is
\begin{equation}
   H^{(1,1)}
  = - \frac{\hbar^2}{2m} \left( \nabla_1^2 + \nabla_2^2 \right)
       + \frac{4\pi \hbar^2a}{m} \delta^3(\bold{r}) \frac{\partial}{\partial r}r.
\end{equation}
Here we introduce the center-of-mass and relative coordinates:
$\bold{R}=(\bold{r}_1+\bold{r}_2)/2$,
$\bold{r}=\bold{r}_1-\bold{r}_2$,
and its absolute value:
$r:=\left|\bold{r}\right|$.
The Schr\"odinger equation for the relative motion is
\begin{equation}
\left( - \frac{\hbar^2}{m} \nabla^2    + \frac{4\pi \hbar^2a}{m} \delta^3(\bold{r}) \frac{\partial}{\partial r}r \right) \psi(\bold{r})=
E \psi(\bold{r}).
\end{equation}
The solutions to this equation are continuous scattering states $\psi_{\rm sc}(r)$ in Eq.~(\ref{eq:def:scat-w.f.}) with energy $E_{\rm sc} = \hbar^2k_{\rm sc}^2/m$
and one bound state $\psi_{b}(r)$ in Eq.~(\ref{eq:def:bound-w.f.}) with the binding energy $E_b = - \hbar^2/(ma^2)$.
The same results can be obtained from the Bethe-Peierls boundary condition (\ref{eq:B-P}).

Using the general formula for the coordinate representation of  the two-particle cluster function, we obtain
\begin{equation}
\langle\bold{r}_1', \bold{r}_2'|U^{(1,1)}_{\infty}|\bold{r}_1, \bold{r}_2\rangle
 =  \frac{\sqrt{8}}{\lambda^3} \,
 e^{- m( \bold{R} - \bold{R}' )^2 / (\beta \hbar^2) }
      \cdot \langle\bold{r}'| u^{({\rm rel})}_{\infty}  |\bold{r}\rangle,
\label{eq:U2-infry}
\end{equation}
where
\begin{equation}
\begin{split}
 & \langle\bold{r}'|u^{({\rm rel})}_{\infty}|\bold{r}\rangle 
  = \theta (a) \, \psi_{b}^*(r)\psi_{b}(r')e^{\beta\hbar^2/(ma^2)}  \\  
 &   +\int_0^\infty \!\!\! dk_{\rm sc} \, e^{-\beta \hbar^2k_{\rm sc}^2/m} 
    \biggl[ \psi_{\rm sc}^*(r)\psi_{\rm sc}(r')
  - \frac{\sin (k_{\rm sc}r) \sin (k_{\rm sc}r')}{2\pi^2rr'} 
    \biggr].
\label{eq:pseudo_real-sp.1}
\end{split}
\end{equation}
Here, $\theta \left( x \right) =  \left(  1+ x/ \left| x \right| \right)/2$. 
As shown in Appendix~\ref{app:U2_pseudo}-2, the integration over $k_{\rm sc}$ gives
%
%
\begin{equation}
\begin{split}
 & \langle\bold{r}'|u^{({\rm rel})}_{\infty}|\bold{r}\rangle 
  =  \frac{1}{4\pi rr'\lambda}
      \Biggl[ \sqrt{2}\,
      e^{ -m(r+r')^2/(4\beta\hbar^2)} \\
 & \, +   \frac{\lambda}{a} \, e^{\beta\hbar^2/(ma^2)}  e^{-(r+r')/a} 
   \operatorname{erfc} \left( \frac{r+r'}{2\hbar}\sqrt{\frac{m}{\beta}} - \frac{\hbar}{a}\sqrt{\frac{\beta}{m}} \right)
 \Biggr].
\end{split}
\label{eq:pseudo_real-sp.2}
\end{equation}
Here, the complementary error function $\operatorname{erfc} \left(x \right)$ is defined in Eq.~(\ref{eq:def:erfc}).

The momentum representation is defined by
\begin{equation}
\begin{split}
 & \langle\bold{k}_1', \bold{k}_2'| U^{(1,1)}_{\infty}|\bold{k}_1, \bold{k}_2\rangle
  =  \frac{1}{(8\pi^3)^2} \!
   \int \!\! d^3 \bold{r}_1 d^3 \bold{r}'_1 d^3 \bold{r}_2 d^3 \bold{r}'_2 \\
 & \qquad\quad \times e^{i\sum_{\alpha=1}^2(\bold{k}'_\alpha\cdot\bold{r}'_\alpha -\bold{k}_\alpha\cdot\bold{r}_\alpha)} 
    \langle\bold{r}_1', \bold{r}_2'|U^{(1,1)}_{\infty}|\bold{r}_1, \bold{r}_2\rangle ,
\end{split}
\label{eq:def:U2_k}
\end{equation}
where
$\bold{k}:=(\bold{k}_1-\bold{k}_2)/2$, $\bold{K}:=\bold{k}_1+\bold{k}_2$
and $k:=\left|\bold{k}\right|$.
We write
\begin{equation}
 \langle\bold{k}_1', \bold{k}_2'|U^{(1,1)}_{\infty}|\bold{k}_1, \bold{k}_2\rangle
  \equiv \delta^3(\bold{K}-\bold{K}') \, \langle\bold{k}_1', \bold{k}_2'|u^{(1,1)}|\bold{k}_1, \bold{k}_2\rangle.
\label{eq:def:u2_k}
\end{equation}
The function $u^{(1,1)}$ is defined only for the case with $\bold{k}_1'+\bold{k}_2' = \bold{k}_1+\bold{k}_2$
and is independent of the volume.
The cluster function defined for a finite volume $V$ is
\begin{equation}
 \langle\bold{k}_1', \bold{k}_2'|U^{(1,1)}_{V}|\bold{k}_1, \bold{k}_2\rangle
 = \frac{8\pi^3}{V} \delta_{\bold{K},\bold{K}'} \,
     \langle\bold{k}_1', \bold{k}_2'|u^{(1,1)}|\bold{k}_1, \bold{k}_2\rangle.
\end{equation}

As shown in Appendix~\ref{app:U2_pseudo}-3, we finally obtain
\begin{equation}
 \langle\bold{k}_1', \bold{k}_2'|u^{(1,1)}|\bold{k}_1, \bold{k}_2\rangle
 =  e^{- \beta\hbar^2\bold{K}^2/(4m)} \, 
      \langle\bold{k}'| u^{({\rm rel})}  |\bold{k}\rangle,
\label{eq:def:u-rel_k}
\end{equation}
where
\begin{equation}
  \langle\bold{k}'|u^{({\rm rel})}|\bold{k}\rangle  =
\begin{cases}
 \displaystyle \frac{\lambda}{2^{3/2}\pi^{5/2}} \, \frac{s(k') - s(k)}{k'^2-k^2}, 
& \text{for} \,\, k\not=k' ;\\
 \displaystyle \frac{\lambda}{(2\pi)^{5/2}} \, \frac{1}{k} \frac{\partial}{\partial k} s(k), 
& \text{for} \,\, k=k' ,\\ 
\end{cases}
\label{eq:u-rel_k}
\end{equation}
\begin{equation}
\begin{split}
s(k)
 =  \frac{\sqrt{m}a}{\sqrt{\beta}\hbar} \frac{1}{1+(ka)^2} 
&   \Biggl[
     e^{-\beta \hbar^2k^2/m}
     -  \frac{2ka}{\sqrt{\pi}} F \left(\frac{\sqrt{\beta} \hbar k}{\sqrt{m}} \right) \\
&       - e^{\beta\hbar^2/(ma^2)}
      \operatorname{erfc} \left( - \frac{\sqrt{\beta}\hbar}{\sqrt{m}a}  \right)
      \Biggr],
\label{eq:gk}
\end{split}
\end{equation}
which gives Eq.~(\ref{eq:def:sx}).

The two-particle cluster function $U^{(2)}_{\infty}$ for the $s$-wave pseudopotential has been given in Ref.~\cite{OU06}
for positive $a$,
but the result in this Appendix holds for arbitrary $a$.

\subsection{Derivation of Eq.~(\ref{eq:pseudo_real-sp.2})}

To derive Eq.~(\ref{eq:pseudo_real-sp.2}), we rewrite Eq.~(\ref{eq:pseudo_real-sp.1}) by using the trigonometric addition and subtraction formulas as
\begin{equation}
\begin{split}
  \langle\bold{r}'|u^{({\rm rel})}_{\infty} |\bold{r} & \rangle 
   = \theta (w) \cdot \psi_{b}^*(r)\psi_{b}(r')e^{w^2}  \\  
 & +
\frac{1}{2\pi^2rr'} 
\frac{ \sqrt{2\pi}}{\lambda}
   \int_0^\infty \!\!\! dx  
 \, e^{-x^2}  \Bigl[  \cos (x \mathcal{R}) \\
& -
 \frac{w}{x^2+w^2} 
    \left( w \cos (x \mathcal{R}) + x \sin (x \mathcal{R})  \right) \Bigr],
\label{eq:u-rel-Integration1}
\end{split}
\end{equation}
where we introduce the dimensionless variables
$x:=\sqrt{\beta \hbar^2\bold{k}_{\rm sc}^2/m}= \lambda \left|\bold{k}_{\rm sc} \right|/\sqrt{2\pi}$,
$w:=\sqrt{\beta}\hbar/(\sqrt{m}a)=\lambda/(\sqrt{2\pi}a)$, and
$\mathcal{R}:=\sqrt{2\pi}(r+r')/\lambda$.
The first term in the integration in Eq.~(\ref{eq:u-rel-Integration1}) gives
\begin{equation}
\int_0^\infty \!\!\! dx  
 \, e^{-x^2} \cos (x \mathcal{R}) = \frac{\sqrt{\pi}}{2}   e^{ -\mathcal{R}^2/4}.
 \label{eq:u-rel-Integration-partial1}
 \end{equation}
The remaining terms are rewritten as
\begin{equation}
\begin{split}
& \int_0^\infty dx  
 \, e^{-x^2}  \frac{1}{x^2+w^2} 
    \bigl( w \cos (x \mathcal{R}) + x \sin (x \mathcal{R})  \bigr) \\
& =  \frac{1}{2} \, \text{Im}  \int_{-\infty}^\infty  dx  
 \,  e^{-x^2}  \frac{e^{i x \mathcal{R}}}{x-iw} \\
& = \frac{1}{2} \, e^{w^2-\mathcal{R}w} 
\left[ \pi \operatorname{sgn} w  +  \int_{-\infty}^\infty \! dx  
 \,    \frac{e^{-x^2}\!}{x}  \sin \bigl(  x (\mathcal{R} -2w) \bigr)
 \right], \\
& = \frac{\pi}{2} \, e^{w^2-\mathcal{R}w} 
\left( \operatorname{sgn} w
 +  \operatorname{erf}  (  \mathcal{R}/2 -w )
 \right),
\end{split}
\label{eq:u-rel-Integration-partial2}
\end{equation}
%
%
where we introduce the error function
$
\operatorname{erf} \left(x \right)
= \left(2 / \sqrt{\pi}\right) \int_{0}^{x} \! dt \, e^{-t^2}$
and the sign function
\begin{equation}
\operatorname{sgn} w =
\begin{cases}
\,\,1 ,  & {\mathrm{for}} \,\,\, w>0 , \\
\,\,0 ,  & {\mathrm{for}} \,\,\, w=0 , \\
-1,  & {\mathrm{for}} \,\,\, w<0 .
\end{cases}
\end{equation}
Substituting Eqs.~(\ref{eq:u-rel-Integration-partial1}) and (\ref{eq:u-rel-Integration-partial2}) into Eq.~(\ref{eq:u-rel-Integration1}), we obtain
\begin{equation}
\begin{split}
  \langle\bold{r}'| & u^{({\rm rel})}_{\infty} |\bold{r}\rangle 
  =   \frac{\sqrt{2}}{4\pi rr'\lambda} \\
& \times      \Bigl[ e^{ - \mathcal{R}^2/4} 
       +   \sqrt{\pi} \, w\, e^{w^2 - \mathcal{R}w} 
  \operatorname{erfc} \left(  \mathcal{R}/2 -w \right) 
 \Bigr],
\end{split}
\label{eq:Derivation-of-Eq.(C5)-last}
\end{equation}
which gives Eq.~(\ref{eq:pseudo_real-sp.2}).

\subsection{Derivation of Eq.~(\ref{eq:u-rel_k})}

To derive Eq.~(\ref{eq:u-rel_k}), 
we substitute Eqs.~(\ref{eq:U2-infry}), (\ref{eq:def:u2_k}), and (\ref{eq:def:u-rel_k})
into Eq.~(\ref{eq:def:U2_k}), and obtain
\begin{equation}
\begin{split}
 &   \langle\bold{k}'|u^{({\rm rel})}|\bold{k}\rangle 
  =  \frac{1}{8\pi^3} \!
   \int \! d^3 \bold{r} \, d^3 \bold{r}' 
  e^{i(\bold{k}'\cdot\bold{r}' -\bold{k}\cdot\bold{r})} 
   \langle\bold{r}'|u^{({\rm rel})}_{\infty}|\bold{r}\rangle .
\end{split}
\label{eq:Derivation-of-Eq.(C10)-1}
\end{equation}
To calculate the Fourier transformation (\ref{eq:Derivation-of-Eq.(C10)-1}),
the following lemma is useful:\\

\noindent
{\it Lemma}.---
\begin{equation}
\begin{split}
& \int \! d^3 \bold{r} \int \! d^3 \bold{r}' \, 
 e^{i ( \bold{k}' \cdot \bold{r}' - \bold{k} \cdot \bold{r}  )} \, 
 \frac{f(r+r')}{rr'} \\
& =  \frac{-16\pi^2}{k'^2 - k^2} \int_0^{\infty} \! d X
  \left( \frac{1}{k'} \sin (k'X) - \frac{1}{k} \sin (kX) \right)  f(X),
\end{split}
\end{equation}
where $r=\left|\bold{r}\right|$, $r'=\left|\bold{r}'\right|$, $k=\left|\bold{k}\right|$, $k'=\left|\bold{k}'\right|$, and $X = r+r'$.
\\

\noindent
{\it Proof.}---
Performing the integration in the spherical coordinates, we obtain
%
\begin{equation}
\begin{split}
& \int \! d^3 \bold{r} \int \! d^3 \bold{r}' \, 
 e^{i ( \bold{k}' \cdot \bold{r}' - \bold{k} \cdot \bold{r}  )} \, 
 \frac{f(r+r')}{rr'} \\
& = 4 \pi^2 \int_0^{\infty} \! d r \int_0^{\infty} \! d r' 
  \int_0^{\pi} \! d \theta \int_0^{\pi} \! d \theta' \, \sin \theta \sin \theta'
 \\
& \qquad \times e^{i ( k'r'\cos\theta' - kr\cos\theta  )} rr' f(r+r') \\ 
& =  \frac{16\pi^2}{kk'} \int_0^{\infty} \! d r \int_0^{\infty} \! d r' \,
  \sin(kr) \sin(k'r')  f(r+r').
\end{split}
\end{equation}
By introducing the new variables $X:= r + r'$ and $Y:= (r' - r) /2$,
we obtain
\begin{equation}
\begin{split}
&   \int_0^{\infty} \! d r \int_0^{\infty} \! d r' \,
  \sin(kr) \sin(k'r')  f(r+r') \\
& =   \int_0^{\infty} \! d X \, f(X) \int_{-X/2}^{X/2} d Y \\
& \qquad \times
  \sin \left( k \left( \textstyle\frac{1}{2}X-Y  \right) \right)
  \sin \left( k' \left( \textstyle\frac{1}{2}X+Y  \right) \right).
\end{split}
\label{eq:dX-dY-integration}
\end{equation}
Integrating this over $Y$, we obtain the lemma.
(Q.E.D.)\\



By Substituting Eq.~(\ref{eq:Derivation-of-Eq.(C5)-last}) into Eq.~(\ref{eq:Derivation-of-Eq.(C10)-1}) and using the above lemma, we obtain
\begin{equation}
  \langle\bold{k}'|u^{({\rm rel})}|\bold{k}\rangle 
  = \frac{\lambda^3}{2^{5/2}\pi^{7/2}} \, \frac{s(x', w) -  s(x, w)}{x'^2-x^2},
\end{equation}
where
\begin{equation}
\begin{split}
 s(x, w)
& =   - \frac{1}{x} \int_0^\infty \! d \mathcal{R} \, \sin (\mathcal{R} x)  
   \biggl( \frac{1}{\sqrt{\pi}} \, e^{ - \mathcal{R}^2/4}  \\
&   \qquad + w\, e^{w^2 - \mathcal{R}w} 
   \operatorname{erfc} \left(  \mathcal{R}/2 -w \right)
 \biggr)  \\
& =  \frac{1}{x^2+w^2}
   \biggl(
     w \, e^{-x^2}
     -  \frac{2}{\sqrt{\pi}} \, x F \left(x \right)  \\
&  \qquad  -  w \, e^{w^2} \operatorname{erfc}  \left( -w \right)
    \biggr).
\label{eq:sx-in-proof}
\end{split}
\end{equation}
Here, we introduce the dimensionless variables
$x:= k \lambda /\sqrt{2\pi}$ and
$x':= k' \lambda /\sqrt{2\pi}$.
This completes the derivation of Eq.~(\ref{eq:u-rel_k}).



\section{Derivation of Eq.~(\ref{BCS-limit_InfSum}) }
\label{app:Weak coupling limit}

\subsection{Derivation 1: Tan's $\Lambda$ function method}

In this Appendix, we derive Eq.~(\ref{BCS-limit_InfSum}) from Eq.~(\ref{eq:XiLad-Irr-1st}).
We first establish the following lemma:\\

\noindent
{\it Lemma}.---
If $ n_F( \bold{k} ) =   ( 1+ z^{-1} e^{\beta \epsilon_{\bold{k}}} )^{-1}$,
then
\begin{equation}
\begin{split}
& \prod_{i=1}^n
   n_F (\bold{k}_{2i-1}) n_F (\bold{k}_{2i}) 
 \int_0^\beta \! d\tau_i \,
 e^{\tau_i (\epsilon_{\bold{k}_{2i+1}} + \epsilon_{\bold{k}_{2i+2}}
 - \epsilon_{\bold{k}_{2i-1}} - \epsilon_{\bold{k}_{2i}}
 )} \\
& = 
 \sum_{l \in \mathbb{Z}}  \prod_{i=1}^n
 \frac{1-n_F(\bold{k}_{2i-1})-n_F(\bold{k}_{2i})}{i\Omega_l -
 (\epsilon_{\bold{k}_{2i-1}} + \epsilon_{\bold{k}_{2i}} -2\mu)},
\label{eq:Lemma}
\end{split}
\end{equation}
where $\Omega_l = 2\pi l /\beta$ 
and the summation $\sum_{l \in \mathbb{Z}}$ extends over all integers $l \in \{ 0, \pm 1, \pm 2, \dots \}$.
\\

\noindent
{\it Proof.}---
The left-hand side of Eq.~(\ref{eq:Lemma}) is rewritten as
\begin{equation}
\begin{split}
& \prod_{i=1}^n
   n_F (\bold{k}_{2i-1}) n_F (\bold{k}_{2i}) 
 \int_0^\beta \! d\tau_i \,
 e^{(\tau_{i-1} - \tau_i ) (\epsilon_{\bold{k}_{2i-1}} + \epsilon_{\bold{k}_{2i}} -2\mu)}.
\label{eq:LemmaProof0}
\end{split}
\end{equation}
We notice the following identities:
\begin{equation}
n_F( \bold{k} )\,e^{\tau ( \epsilon_{\bold{k}} -\mu )}
 =
 \frac{1}{\beta} \sum_{n \in \mathbb{Z}}\frac{e^{i\omega_n \tau}}{i\omega_n -( \epsilon_{\bold{k}} -\mu )},
\label{eq:LemmaProof1}
\end{equation}
where $\omega_n = \frac{2\pi}{\beta}(n +\frac12)$
and the sum $\sum_{n \in \mathbb{Z}}$ runs over all integers $n \in \{ 0, \pm 1, \pm 2, \dots \}$;
\begin{equation}
\begin{split}
& \frac{1}{\beta^n} \int_0^\beta\!d\tau_1\cdots\int_0^\beta\!d\tau_n
 \prod_{i=1}^ne^{i(\tau_{i-1}-\tau_i)(\omega_{2i-1} + \omega_{2i} )} \\
& = \prod_{i=1}^n\,\delta_{\omega_{2i+1} + \omega_{2i+2} - \omega_{2i-1} - \omega_{2i}} ;
\end{split}
\end{equation}
and 
\begin{equation}
\begin{split}
& \sum_{n_1 \in \mathbb{Z}} \, \sum_{n_2 \in \mathbb{Z}} \,
 \frac{1}{i\omega_{n_1} -( \epsilon_{\bold{k}_1} -\mu )} \cdot
 \frac{1}{i\omega_{n_2} -( \epsilon_{\bold{k}_2} -\mu )} \\
& = (-\beta) \sum_{l \in \mathbb{Z}}
 \frac{1-n_F(\bold{k}_{1})-n_F(\bold{k}_{2})}{i\Omega_l -
 (\epsilon_{\bold{k}_{1}} + \epsilon_{\bold{k}_{2}} -2\mu)},
 \label{eq:LemmaProof3}
\end{split}
\end{equation}
where $\Omega_l = 2\pi l /\beta$. 
Combining the above identities (\ref{eq:LemmaProof1})-(\ref{eq:LemmaProof3}) with (\ref{eq:LemmaProof0}), we obtain the lemma.
(Q.E.D.)\\

Substituting the above Lemma with 
$\hbar^2(\bold{p}_{i+1}^2 - \bold{p}_i^2 )/m
= \epsilon_{\bold{k}_{2i+1}} + \epsilon_{\bold{k}_{2i+2}}
 - \epsilon_{\bold{k}_{2i-1}} - \epsilon_{\bold{k}_{2i}}$,
$\bold{k}_{2i-1}= (1/2)\bold{K} -\bold{p}_i$,
and $\bold{k}_{2i}= (1/2)\bold{K} +\bold{p}_i$
into Eq.~(\ref{eq:XiLad-Irr-1st}),
we obtain
\begin{equation}
\begin{split}
& \mathcal{P}_{{\rm pair}}^{{\rm 1st}}
 = \sum_{l \in \mathbb{Z}} \sum_{\bold{K}}
 \sum_{n=1}^\infty \frac{1}{n}  \left( \frac{4\pi \hbar^2 a}{Vm} \right)^n 
    \sum_{\bold{p}_1,\dots,\bold{p}_{n}}     \\
& \quad \prod_{i=1}^n
  \frac{1-n_F(\epsilon_{\frac{\bold{K}}{2}+\bold{p}_i})-n_F(\epsilon_{\frac{\bold{K}}{2}-\bold{p}_i})}
 {i\Omega_l - (\epsilon_{\frac{\bold{K}}{2}+\bold{p}_i}+\epsilon_{\frac{\bold{K}}{2}-\bold{p}_i} -2\mu)}
  \Lambda \left(\bold{p}_i \right) \\
& = \sum_{l \in \mathbb{Z}} \sum_{\bold{K}}
 \sum_{n=1}^\infty \frac{1}{n} 
      \\
& \quad \times
 \left[
  \frac{4\pi \hbar^2 a}{Vm} \sum_{\bold{p}} 
    \frac{1-n_F(\epsilon_{\frac{\bold{K}}{2}+\bold{p}})-n_F(\epsilon_{\frac{\bold{K}}{2}-\bold{p}})}
 {i\Omega_l - (\epsilon_{\frac{\bold{K}}{2}+\bold{p}}+\epsilon_{\frac{\bold{K}}{2}-\bold{p}} -2\mu)}
  \Lambda \left(\bold{p} \right) 
  \right]^n .
\label{eq:App-BCSlimit-Last1}
\end{split}
\end{equation}
Using the property of Tan's $\Lambda$ function (\ref{eq:Lambda-fn-prop}),
we have
\begin{equation}
\begin{split}
& \sum_{\bold{p}} 
    \frac{1-n_F(\epsilon_{\frac{\bold{K}}{2}+\bold{p}})-n_F(\epsilon_{\frac{\bold{K}}{2}-\bold{p}})}
 {i\Omega_l - (\epsilon_{\frac{\bold{K}}{2}+\bold{p}}+\epsilon_{\frac{\bold{K}}{2}-\bold{p}} -2\mu)}
  \Lambda \left(\bold{p} \right) \\
& =  
   \sum_{\bold{p}}
  \left[ 
  \frac{1-n_F(\epsilon_{\frac{\bold{K}}{2}+\bold{p}})-n_F(\epsilon_{\frac{\bold{K}}{2}-\bold{p}})}
 {i\Omega_l - (\epsilon_{\frac{\bold{K}}{2}+\bold{p}}+\epsilon_{\frac{\bold{K}}{2}-\bold{p}} -2\mu)}
  + \frac{m}{\hbar^2\bold{p}^2} 
\right]
 \Lambda \left(\bold{p} \right)  \\
& =  
   \sum_{\bold{p}}
  \left[ 
  \frac{1-n_F(\epsilon_{\frac{\bold{K}}{2}+\bold{p}})-n_F(\epsilon_{\frac{\bold{K}}{2}-\bold{p}})}
 {i\Omega_l - (\epsilon_{\frac{\bold{K}}{2}+\bold{p}}+\epsilon_{\frac{\bold{K}}{2}-\bold{p}} -2\mu)}
  + \frac{m}{\hbar^2\bold{p}^2} 
\right] .
\label{eq:App-BCSlimit-Last2}
\end{split}
\end{equation}
By combining Eqs.~(\ref{eq:App-BCSlimit-Last1}) and (\ref{eq:App-BCSlimit-Last2}),
Eq.~(\ref{BCS-limit_InfSum}) follows immediately.

\subsection{Derivation 2: standard regularization method}

Here, we derive Eq.~(\ref{BCS-limit_InfSum}) by using the standard regularization method.
We consider a delta-function potential
\begin{equation}
v \left(\left|\bold{r}_i-\bold{r}_j \right| \right) 
= g \, \delta \left(\left|\bold{r}_i-\bold{r}_j \right| \right).
\label{eq:contact_intaraction}
\end{equation}
Here, $g\, (<0)$ is related to the $s$-wave scattering length $a$ as
\begin{equation}
\frac{m}{4\pi \hbar^2 a}
= \frac{1}{g} + \frac{1}{V} \sum_{\bold{k}} \frac{1}{2 \epsilon_{\bold{k}}},
\label{eq:regularization}
\end{equation}
where $\epsilon_{\bold{k}} = \hbar^2 k^2/(2m)$. (See e.g. Ref.~\cite{MRE93}).
Corresponding to Eq.~(\ref{eq:pseudopotential_momentum}),
we obtain
$\langle\bold{k}_1'; \bold{k}_2'| \,v\, |\bold{k}_1; \bold{k}_2\rangle 
= (g/V) \, \delta_{\bold{K},\bold{K}'}$,
where $\bold{K}=\bold{k}_{1}+\bold{k}_{2}$ and $\bold{K}'=\bold{k}_{1}'+\bold{k}_{2}'$.


Following the procedure of deriving Eq.~(\ref{eq:XiLad-Irr-1st}) from Eq.~(\ref{eq:1st-order_of_U2}) in Sec. IV-C and
by replacing the factor $(4\pi a/m) \Lambda \left(\bold{p}_i \right)$ with $g$,
we obtain
\begin{equation}
\begin{split}
 \mathcal{P}_{{\rm pair}}^{{\rm 1st}}
 = & \sum_{\bold{K}}
 \sum_{n=1}^\infty \frac{1}{n}  \left( \frac{g}{V} \right)^n 
    \sum_{\bold{p}_1,\dots,\bold{p}_{n}}  
  \prod_{i=1}^n
  n_F \left( {\textstyle \frac{1}{2}} \bold{K} + \bold{p}_i \right)  \\
& \times  n_F \left( {\textstyle \frac{1}{2}} \bold{K} - \bold{p}_i \right) 
   \int_0^\beta \!\! d\tau_i \,
 e^{\tau_i \hbar^2(\bold{p}_{i+1}^2 - \bold{p}_i^2 )/m}.
\label{eq:XiLad-Irr-1st-g}
\end{split}
\end{equation}
Substituting 
$\hbar^2(\bold{p}_{i+1}^2 - \bold{p}_i^2 )/m
= \epsilon_{\bold{k}_{2i+1}} + \epsilon_{\bold{k}_{2i+2}}
 - \epsilon_{\bold{k}_{2i-1}} - \epsilon_{\bold{k}_{2i}}$,
$\bold{k}_{2i-1}= (1/2)\bold{K} -\bold{p}_i$,
and $\bold{k}_{2i}= (1/2)\bold{K} +\bold{p}_i$
into Eq.~(\ref{eq:XiLad-Irr-1st-g})
with the Lemma of the previous subsection,
we obtain
\begin{equation}
\begin{split}
 \mathcal{P}_{{\rm pair}}^{{\rm 1st}}
& = \sum_{l \in \mathbb{Z}} \sum_{\bold{K}}
 \sum_{n=1}^\infty \frac{1}{n}   \\
&  \times \left[
  \frac{g}{V} \sum_{\bold{p}} 
    \frac{1-n_F(\epsilon_{\frac{\bold{K}}{2}+\bold{p}})-n_F(\epsilon_{\frac{\bold{K}}{2}-\bold{p}})}
 {i\Omega_l - (\epsilon_{\frac{\bold{K}}{2}+\bold{p}}+\epsilon_{\frac{\bold{K}}{2}-\bold{p}} -2\mu)}
  \right]^n .
\label{eq:App-BCSlimit-Last-g}
\end{split}
\end{equation}
By substituting Eq.~(\ref{eq:regularization}) into Eq.~(\ref{eq:App-BCSlimit-Last-g}),
we obtain Eq.~(\ref{BCS-limit_InfSum}).

\newpage \quad \\
\quad \\
\quad \\
\quad \\
\quad \\
\quad \\
\quad \\

\bibliographystyle{apsrev}


\end{document}